\documentclass[11pt]{article}
\usepackage{latexsym,array,theorem,mathrsfs,bm,float}
\usepackage{psfrag}
\usepackage{amsfonts,amsmath,amssymb,latexsym,array,afterpage,
theorem,mathrsfs,bm,float,epsfig,color,tabularx,graphicx,
here,multirow,mediabb}
\usepackage[dvipdfmx]{hyperref}
\usepackage{fancybox}

  \makeatletter
    
    \@addtoreset{equation}{section}
  \makeatother

\usepackage[top=30mm,bottom=35mm,left=25mm,right=25mm]{geometry}

\newcommand{\bea}{\begin{eqnarray}}
\newcommand{\ena}{\end{eqnarray}}
\newcommand{\beann}{\begin{eqnarray*}}
\newcommand{\enann}{\end{eqnarray*}}

\newcommand{\ma}[1]{\mbox{$\mathcal{#1}$}}

\newcommand{\ti}{\tilde}

\newcommand{\calhR}[1]{\raisebox{2ex}{\tiny ({\em h})}\hspace{-0.8em}{\ma R}}
\newcommand{\pd}{\partial}
\newcommand{\BS}{\boldsymbol}
\newcommand{\MC}{\mathcal}

\newcommand{\MB}{\mathbb}

\newcommand{\p}{\prime}

\newcommand{\acknowledgemnet}{\section*{Acknowledgement}}

\newif\iffigure
\figurefalse
\figuretrue

\begin{document}

\title{\LARGE{\bf{
Many phases in a hairy box in three dimensions}}
}

\author{
\\
Shoichiro Miyashita\thanks{e-mail address : s-miyashita"at"gms.ndhu.edu.tw} 
\\ \\ 
{\it Department of Physics, National Dong Hwa University, Hualien, Taiwan, R.O.C.} 
\\ \\
}

\date{~}


\maketitle
\begin{abstract}
In this paper, I investigate gravitational thermodynamics of the Einstein-Maxwell-scalar system in three dimensions without a cosmological constant. In the previous work by Krishnan, Shekhar, and Bala Subramanian (Nucl. Phys. B \textbf{958} (2020) 115115. \cite{KSB2020}), it was argued that this system has no BH saddles, but has only empty (flat space)  saddles and boson star saddles. 
It was then concluded that the structure of the thermodynamic phase space is much simpler than in the higher dimensional cases.
I will show that, in addition to the known boson star and empty saddles, three more types of saddles exist in this system: the BG saddle, its hairy generalization, and a novel configuration called the boson star-PL saddle. 
As a result, the structure is richer than one might naively expect and is  very similar to the higher dimensional ones. 
\end{abstract}

\clearpage

\clearpage

\section{Introduction}

The work of Hawking and Page \cite{HP1983} offered the possibility that pure gravity may thermalize and showed that the thermal equilibrium may be described by the Euclidean gravitational path integral(GPI)
\footnote{
In general, GPI has several problems, such as the non-renormalizability problem and the integration contour problem \cite{GHP1978}. However, there is compelling evidence supporting the validity of this approach, including the derivation of ``BH thermodynamics''\cite{GH1976}, the AdS/CFT correspondence \cite{Maldacena1998, GKP1998, Witten1998}, and the derivation of the Page curve of Hawking radiation \cite{AMMZ2020, PSSY2022, AHMST2020}. In this paper, I will assume that GPI can capture the thermodynamic properties of quantum gravity.
}
when the spacetime has anti de Sitter(AdS) boundaries. Later, York\cite{York1986} showed that this is also true for the Dirichlet boundary, where the area of the boundary sphere is set to finite. In both cases, black holes(BH) play the role of the high temperature phase, corresponding to the dominance of Euclidean BH saddles in the GPI for the partition function. This fact also seems to hold when some matter fields couple to gravity and the boundary is the AdS one. Some examples are Einstein-Maxwell(EM)\cite{CEJM1999}, Einstein-Maxwell-scalar(EMS)\cite{HHH20082, MKF2010, BBBLMU2010, BKB20161, AMS2016}, Einstein-Yang-Mills\cite{BH20001, BH20002, Winstanley1999, MRT2006, KKRS2015}, Einstein-Yang-Mills-Higgs\cite{LMS2000, LS1999, LMS20101, LMS20102, GL2015, MM2016}. On the other hand, when the boundary is a Dirichlet one, the situation seems to be different. A series of investigations of EM thermodynamics with Dirichlet boundary\cite{BBWY1990, BKB20162, Miyashita20241, Miyashita20242} showed that the high temperature phase is not described by BHs, but by, what I call, BGs.
\footnote{
BG stands for bag of gold. For the reason (which is not important), see \cite{Miyashita2021}. 
}
BGs are geometries with horizon, similar to BHs, but whose area is larger than that of the Dirichlet boundary. Obviously, this type of geometries does not appear when the boundary is AdS one, since the boundary area is infinite and its horizon area cannot be larger. BGs are relevant to the Dirichlet boundary case. And they are important for a complete thermodynamic description for this case. 

The main purpose of this paper is to investigate the existence of ``hairy'' BGs for a system which is known to have hairy solutions, to construct hairy BGs explicitly, and to discuss thermodynamics and the role of hairy BGs in thermodynamics. In this paper, I focus on the EMS system without a cosmological constant $\Lambda$ in three dimensions;
\bea
I^{E}[\BS{g}, \BS{A}, \varphi]= I^{E}_{bulk}[\BS{g}, \BS{A}, \varphi] + I^{E}_{bdy}[\BS{g}] ~ , \hspace{9.cm} \label{EqSec1Action}  \\
I^{E}_{bulk}[\BS{g}, \BS{A}, \varphi] = \frac{-1}{16\pi G} \int_{\MC{M}} d^3 x \sqrt{g} \MC{R}  + \frac{1}{16 \pi}\int_{\MC{M}} d^3 x \sqrt{g} F_{\mu\nu} F^{\mu\nu} + \int_{\MC{M}} d^3 x \sqrt{g}  g^{\mu\nu} {\rm D}_{\mu} \varphi ({\rm D}_{\nu} \varphi)^{\ast} ~ , \\
I^{E}_{bdy}[\BS{g}] =  \frac{-1}{8\pi G} \int_{\pd\MC{M}} d^2 y \sqrt{\gamma} \Theta ~. \hspace{9.1cm}
\ena
One reason why I chose a $\Lambda=0$ and three dimensional theory is that the analysis would be simpler than in higher dimensions or $\Lambda<0$ since we might not have BH solutions, i.e. the structure of the phase diagram could be simpler. (But as I will explain later, the phase diagram of this system is not as simple as claimed in the previous paper \cite{KSB2020}. It is as complicated as the AdS boundary case, or holographic superconductors.) This would be a good first step in understanding the nature of hairy BGs. Another reason is related to the investigation of the three dimensional EM system with $\Lambda=0$ \cite{Miyashita20242}. In that paper, I found that the specific heat of BGs for that case is not positive but zero at the zero-loop level. Since there are no other saddles in the topology sector, it seems that the system is not thermodynamically stable. 
\footnote{If we assume that both the zero-loop calculation and the restriction to the simple topology sectors are sufficient to extract the thermodynamic property of gravity.
}
Although I suspect that one-loop corrections may eventually resolve this problem, it is not explicitly shown at the moment. Therefore, I would like to check whether this property of BGs is universal for three dimensions without $\Lambda$, or just specific to the BGs in the EM system.

To summarize, the main motivation of this paper is to examine whether the richness of the phase diagram observed in higher-dimensional gravitational systems persists in the simpler setting of three dimensional EMS system with a Dirichlet boundary. In particular, I aim to construct and classify novel saddle geometries such as hairy BG and boson star-PL saddles, and to understand their thermodynamic roles. This investigation may serve as a useful step toward clarifying the universal features of gravitational thermodynamics with a Dirichlet boundary.

The organization of this paper is as follows.

In Section 2, I give the details of the setup, including the equations of motion for static and circularly (``spherically'') symmetric configurations, boundary conditions, and thermodynamic quantities.

In Section 3, I review the thermodynamic properties of the EM system \cite{Miyashita20242}, to which the EMS system reduces when $q=0$
\footnote{
As long as we consider the zero-loop order and with the boundary condition $\varphi(r_{b})=0$.
}
 where $q$ is the coupling constant between the Maxwell field and the scalar field. In this system, there are no BH saddles. Instead, there are BG saddles. I explain what BG saddle is and it is the high temperature phase of the system. 
 
In Section 4, I investigate the thermodynamics of the EMS system with finite $q$. The thermodynamics of this system was previously investigated in \cite{KSB2020}, where it was argued that the system has only two saddles, the empty saddle (flat space saddle) and the boson star saddle. I show that three saddles were missing in their analysis. One is hairless BG saddle, which also exists in the $q=0$ case. Another one is a hairy version of the BG saddle. The third one is what I call ``boson star-PL'' saddle. All of them have no analogues in the more familiar AdS boundary case. I explain the latter two in detail and argue that the resulting phase  diagram is very similar to the AdS cases.

In Section 5, I summarize the results of this paper and list possible future directions.

\section{EOM, boundary conditions, and thermodynamic quantities}

In this paper, I consider grand canonical ensembles in three dimensional EMS system. The partition function $Z(\beta, \mu, r_{b})$ is formally defined by the Euclidean GPI;
\bea
Z(\beta, \mu, r_{b}) = \sum_{\MC{M}} \int_{\MC{M}} \MC{D} \BS{g} \MC{D} \BS{A} \MC{D} \varphi ~ e^{-I^{E}[\BS{g}, \BS{A}, \varphi]} ~ . \label{EqSec2GPI}
\ena
Its Euclidean action is given by (\ref{EqSec1Action}) where the covariant derivative of the scalar field is given by ${\rm D}_{\mu} \varphi = \nabla_{\mu} \varphi - i q A_{\mu} \varphi$. This form of the Euclidean action follows the standard Gibbons–Hawking–York prescription for Dirichlet boundary conditions \cite{York1972, GH1976, York1986, BBWY1990}. There are two parameters in this action, $q$ and $G$. Although I leave $G$ untouched, we can think that this parameter determines the unit. Therefore, the only parameter that controls the theory is $q$. I use the zero-loop approximation of the GPI and also assume that the dominant contribution comes from solutions with high symmetry. Therefore, I focus on static circularly symmetric solutions and use the standard field ansatz, following the previous works, e.g. \cite{KSB2020, HHH20082};
\bea
ds^2 = g_{\mu\nu}dx^{\mu}dx^{\nu} = \Delta(r) f(r) dt^2 + \frac{1}{f(r)}dr^2 + r^2 d\phi^2  ~~~ \label{EqSec2metric} \\
(~ t\in(0, \beta_{0}), ~~~ \phi \in (0, 2\pi) ~ ) ~, \notag \\
A_{\mu}dx^{\mu} = -i a(r)dt ~, \hspace{3.9cm} \label{EqSec2gauge}\\
\varphi = \varphi(r)  ~~~~~~ (\varphi(r) \in \MB{R}) ~.\hspace{2.1cm} 
\ena
Using the variables $\Delta(r), f(r), a(r), \varphi(r)$, the equations of motion(EOM) can be written as
\bea
\Delta'= 32\pi G r \left( \frac{q^2 a^2 \varphi^2}{f^2 } + \Delta (\varphi')^2   \right) ~ , \hspace{3.4cm} \label{EqSec2EOMdelta}  \\
 f' = -16\pi G r \left(  \frac{1}{8\pi} \frac{ (a')^2 }{\Delta } + f (\varphi')^2  + \frac{ q^2 a^2 \varphi^2 }{\Delta f }  \right) ~ , \hspace{1.4cm} \label{EqSec2EOMf}  \\
a'' + \frac{ a'}{ r}   - 16\pi G r a'  \left(  (\varphi')^2  + \frac{q^2 a^2 \varphi^2}{f^2 \Delta }  \right) - \frac{8\pi  q^2 a \varphi^2}{f}  = 0  ~ , \label{EqSec2EOMa}   \\
\varphi'' +  \frac{  \varphi' }{r} - \frac{2 G r (a')^2 \varphi'  }{\Delta f }  + \frac{ q^2 a^2 \varphi}{ \Delta f^2}  = 0 ~. \hspace{3.1cm} \label{EqSec2EOMphi}  
\ena

An ensemble is specified by the inverse temperature $\beta$, the chemical potential $\mu$, and ($\frac{1}{2\pi}$ times) the length of the boundary $r_{b}$. These variables are defined by the boundary values of the metric and the gauge field;
\bea
\beta = \beta_{0} \times \sqrt{g_{tt}}|_{\pd \MC{M}} = \beta_{0} \sqrt{\Delta(r_{b}) f(r_{b})} ~ ,  \label{EqSec2beta} \\ 
\mu = i \left. \frac{A_{t}}{\sqrt{g_{tt}}} \right|_{\pd \MC{M}} = \frac{a(r_{b})}{\sqrt{\Delta(r_{b}) f(r_{b})}} ~ ,\hspace{0.8cm} \label{EqSec2mu} \\
r_{b} = \sqrt{g_{\phi \phi}}|_{\pd \MC{M}} = r_{b} ~ .\hspace{3.0cm}
\ena 
The boundary condition of the GPI for grand canonical ensembles (\ref{EqSec2GPI}) is Dirichlet type and all the above quantities are held fixed. In addition, the boundary value of the scalar field is also fixed under the Dirichlet type boundary condition. I fix $\varphi(r_{b})$ as
\bea
\varphi(r_{b})=0 ~ . \label{EqSec2bdyphi}
\ena 
Using ``gauge fields'' in the bulk, we can define conserved currents and the associated conserved charges on the boundary \cite{BY1993}. In general, when the boundary geometry admits the (normalized) boundary Killing vectors $\xi_{(t)} \propto \pd_{t}, ~ \xi_{(\phi)} \propto \pd_{\phi} $, using the Brown-York tensor \cite{BY1993}
\bea
\tau_{ij} =  \frac{2}{\sqrt{\gamma}} \frac{\delta I^{E}}{\delta g^{ij}} =  \frac{1}{8\pi G}(\Theta_{ij} - \gamma_{ij} \Theta )  ~ ,
\ena
the energy current $j_{(E)}^{i}$ and momentum current $j_{(J)}^{i}$ are defined by
\bea
j_{(E)}^{i} = \tau^{i}_{~ j} \xi_{(t)}^{j} ~ , \hspace{0.3cm} \\
j_{(J)}^{i} =  \frac{1}{i} \tau^{i}_{~ j} \xi_{(\phi)}^{j}  ~ . 
\ena
The charge current is defined by
\bea
j^{i}_{(Q)} = \frac{1}{i}\frac{1}{\sqrt{\gamma}} \frac{\delta I^{E}}{\delta A_{i}} = \frac{1}{4 \pi i} n_{\mu} F^{\mu i} ~ .
\ena
Then, if we consider the form of metric (\ref{EqSec2metric}), conserved charges $\MC{Q} (= E, J, Q)$ is given by
\bea
\MC{Q} = \int_{0}^{2\pi} d\phi \sqrt{\gamma_{\phi\phi}} u_{i} j^{i}_{(\MC{Q})} ~~~~~~ (\MC{Q}=E, J, Q) ~ ,
\ena
where $u_{i}$ is the normal vector to $t={\rm const.}$ surfaces and given by $u_{i} = (u_{t}, u_{\phi})=(\sqrt{\Delta f}, 0) $.
\footnote{
Typically, the aforementioned currents and charges are defined in a  Lorentzian signature. In that case, conservation is with respect to the spacelike hypersurfaces on the boundary. Consider two spacelike hypersurfaces on the boundary, denoted by $\Sigma$, $\Sigma^{\p}$ and their normal vector $u_{i}$. Then, the following equality holds;
\beann
\int_{\Sigma} d^{d-2}z \sqrt{\sigma} u_{i} j^{i}_{(\MC{Q})} = \int_{\Sigma^{\p}} d^{d-2}z \sqrt{\sigma} u_{i} j^{i}_{(\MC{Q})} ~ .
\enann
This is analogous to the standard conservation law in QFT on a fixed curved background. ($z$ is the coordinate on $\Sigma, \Sigma^{\p} $ and $\BS{\sigma}$ is the induced metric of $\Sigma, \Sigma^{\p} $.) See \cite{BY1993} for details.
}
Since I will not consider rotations, $J=0$. In the following sections, there are two classes of saddles, one has $n_{r} >0$ and the other has $n_{r}<0$. Depending on the sign of $n_{r}$, the expressions for the conserved charges are slightly different and are as follows;\\
\underline{$n_{r} > 0$  ~~~ (empty saddle, boson star saddle)}
\bea
E= - \frac{\sqrt{f(r_{b})}}{4G} ~ , \hspace{0.15cm} \\
Q=  \frac{r_{b}}{2}  \frac{a'(r_{b})}{\sqrt{\Delta(r_{b}) } } ~ . 
\ena
\underline{$n_{r} < 0$ ~~~ (BG saddle, hairy BG saddle, boson star-PL saddle)}
\bea
E= \frac{\sqrt{f(r_{b})}}{4G} ~ , \label{EqSec2BGenergy} \hspace{0.75cm} \\
Q =  - \frac{r_{b}}{2}  \frac{a'(r_{b})}{\sqrt{\Delta(r_{b}) } } ~ .
\ena
The free energy at the zero-loop order is given by the on-shell action of the dominant saddle. Here, I abuse the notion of free energy and define the free energy $F_{\rm (A)}$ for a saddle A by 
\bea
\beta F_{\rm (A)}(\beta, \mu, r_{b}) = I^{E}|_{\rm on-shell(A)}(\beta, \mu, r_{b}) ~ . \label{EqSec2onshell}
\ena
For example, as I will review in the next section, in the EM system there are two types of saddles, empty saddle and BG saddle. In this case, ${\rm A} \in \{ {\rm empty, ~ BG} \}$ and we may have two free energies $F_{\rm empty}$ and $F_{\rm BG}$. For a given $(\beta, \mu, r_{b})$, the true free energy $F(\beta, \mu, r_{b})$ of the system is given by $F(\beta, \mu, r_{b}) = \min\{ F_{\rm empty}(\beta, \mu, r_{b}), F_{BG}(\beta, \mu, r_{b}) \}$.

\section{Thermodynamics of $ q=0 $ ~ ($=$ Einstein-Maxwell)}
In this section, I investigate the thermodynamics of the EMS system with $q=0$. In this case, non-trivial scalar configurations are not allowed, and the only allowed configuration is 
\bea
\varphi(r)=0 ~ .
\ena
Then, the EOM (\ref{EqSec2EOMdelta}) - (\ref{EqSec2EOMa}) reduce to those of the EM system
\bea
\Delta'= 0 ~ ,  ~~~~ f' = -2 G r \frac{ (a')^2 }{\Delta } ~, ~~~~ a'' + \frac{ a'}{ r}   = 0 ~. \label{EqSec3EOM}
\ena
Therefore, the thermodynamic properties are the same as in the EM system studied in \cite{Miyashita20242}. Below I review the results of \cite{Miyashita20242} with correcting some errors in that paper. 
\footnote{
In that paper, I made a mistake about the free energy and the energy of BG saddles. The fact that the free energy is a linear function of $T$ is not changed by this error. So the final conclusion of \cite{Miyashita20242} is still true. However, the ``transition temperature'' and the ``phase diagram'' are slightly modified.
}

\subsection{Saddles}
The EOM (\ref{EqSec3EOM}) has two types of solutions, or saddles;
\bea
\underline{\rm empty ~ saddle} ~~~~~~ f(r) = 1~, ~~~ a(r) =\mu \hspace{4.5cm} r \in (0, r_{b}) ~ , \hspace{0.2cm} \label{EqSec3empty} \\
\underline{\rm BG ~ saddle}  ~~~~~~ f(r) = 8 G Q^2 \log \left(\frac{r_{G}}{r} \right)~ , ~~~ a(r) = 2 Q \log \left(\frac{r_{G}}{r} \right) ~~~~~ r \in (r_{b}, r_{G}) ~ , \label{EqSec3BG}
\ena
where $Q$ and $r_{G}$ are functions of the parameters of the ensemble $(\beta, \mu, r_{b})$ that will be derived in the next subsection. In both cases, I set $\Delta(r) = 1$ since it just corresponds to a rescaling of the time coordinate. The former saddle is just a Euclidean flat space inside the box with a constant gauge field so should be familiar to almost all the readers. Much less familiar is the latter, which I call the BG saddle. 
\footnote{
As I will explain shortly, this saddle is similar to, but different from, the BH saddle. That is why I use the other name for this saddle. Sometimes it is called ``cosmological side'' or ``cosmological type'', presumably because this type of saddle was first studied in the context of dS thermodynamics\cite{Miyashita2021, DF2022, BJ2022}, where the bolt corresponds to the cosmological horizon of dS spacetime. 
}
$r_{G}$ is the radial position of the bolt and it is larger than that of the boundary $r=r_{b}$. The $r-\phi$ section and the $r-t$ section of both saddles are shown in Figure \ref{FIG1}. 
                                                 %
\iffigure
\begin{figure}[h]
\begin{center}
	\includegraphics[width=15.cm]{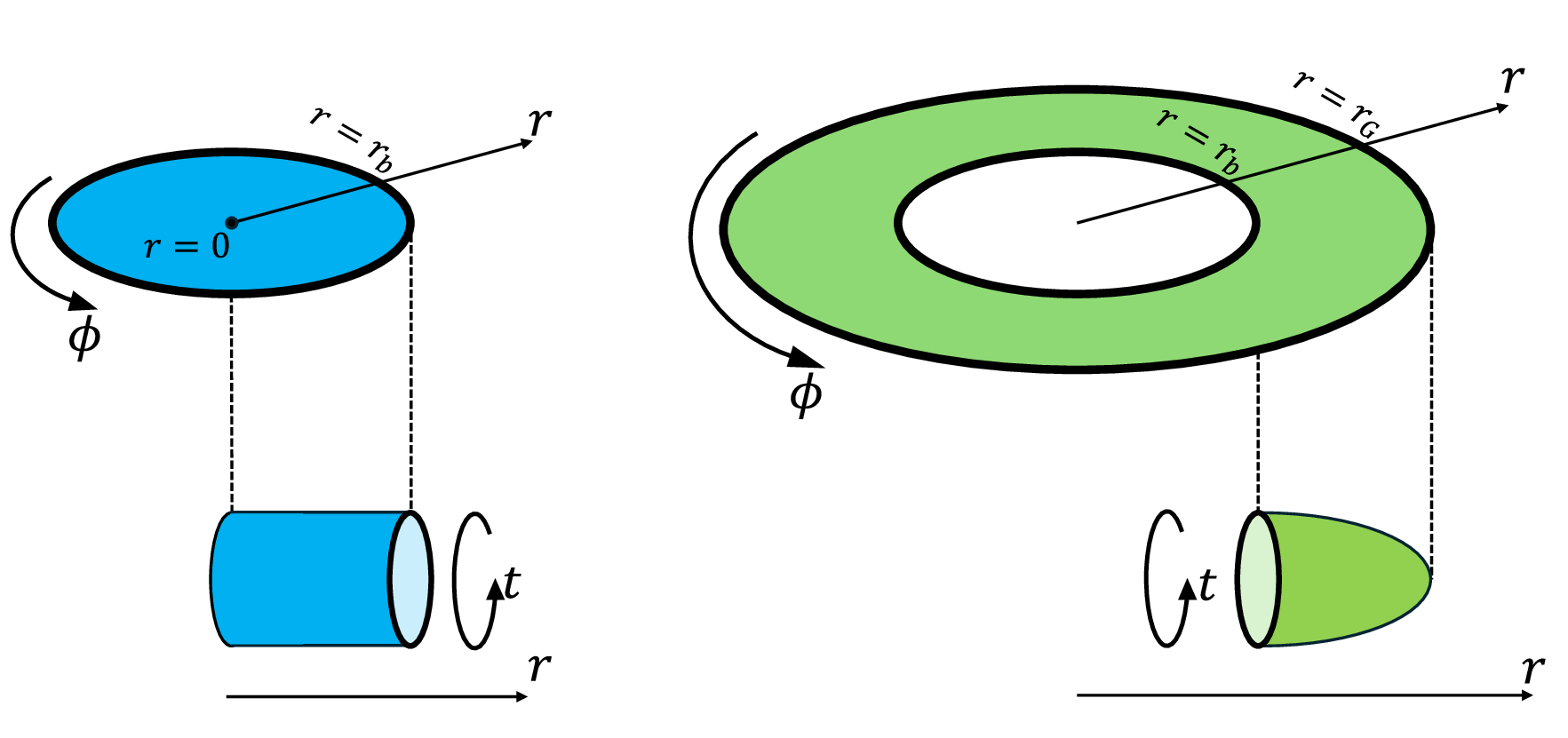} 
	\caption{ Schematic picture of empty saddle (left) and BG saddle (right).  The colored regions represent the bulk of the saddles. The top is the $r-\phi$ section and the bottom is the $r-t$ section. The $r-\phi$ section of the empty saddle has one boundary at $r=r_{b}$. Therefore, readers may easily recognize this geometry as being inside a box. On the other hand, the $r-\phi$ section of the BG saddle has two ``boundaries'' at $r=r_{b}$ and $r= r_{G}$. However, the latter is not a true boundary since the $r-t$ section is capped off, i.e. $r=r_{G}$ is the bolt. The true boundary is only at $r=r_{b}$ for this saddle as well.  This situation is similar to that of BH saddles, which have two boundaries at $r = r_H$ and $r = r_b$ in the t = const. surface. The former is not a true boundary since $r = r_H$ is the bolt. The difference lies in the position of the bolt: $r_{H}< r_{b}$ for BH saddles and $r_{G}>r_{b}$ for BG saddles.  }
\label{FIG1}
\end{center}
\end{figure}
\fi
                                                 %
Unlike the empty saddle, the bulk of the BG saddle appears to be the ``outside'' of the box boundary. However, since $r=r_{G}$ is the bolt, i.e. the $r-t$ section is capped off at $r=r_{G}$, there are no boundaries except $r=r_{b}$. Therefore, the bulk of BG saddles is actually the inside of the boundary. Since these are obviously not BH saddles, they evade the no black hole theorem in three-dimensional gravity \cite{Ida2000, KSB2020} and the system admits finite entropy states even in three dimensions. The existence of this kind of saddles was emphasized by the author and others independently in \cite{Miyashita2021, DF2022, BJ2022} and their importance for gravitational thermodynamics was emphasized by the author in \cite{Miyashita20241}. 

\subsection{Thermodynamics}
Let's focus on the BG saddles. In (\ref{EqSec3BG}), we have the unknown constants $Q$ and $r_{G}$. These are determined as follows. First, squaring the both sides of the equation for $\mu$ (\ref{EqSec2mu}) leads to
\bea
\mu^2 = \frac{a(r_{b})^2}{f(r_{b})} = \frac{4 Q^2 \left( \log \left(\frac{r_{G}}{r_{b}} \right) \right)^2}{ 8 G Q^2 \log \left(\frac{r_{G}}{r_{b}} \right)} = \frac{ \log \left(\frac{r_{G}}{r_{b}} \right) }{ 2 G } \notag \\
\Leftrightarrow ~~ r_{G} = r_{b} e^{2 G \mu^2} ~ . \hspace{2cm}
\ena
Although the spacetime is not asymptotically flat, the regularity at the bolt ($r = r_{G}$) imposes a smoothness condition which determines $\beta_{0}$($=$ the circumference of $t$, not the inverse temperature $\beta$), as in the standard Euclidean approach. It is given by
\bea
\beta_{0} = -\left. \frac{4 \pi}{f'} \right|_{r=r_{G}} = \frac{\pi r_{G}}{2G Q^2} = \frac{\pi r_{b}}{2G Q^2} e^{2 G \mu^2} ~ ,
\ena
and $\sqrt{f(r_{b})}$ can be written as
\bea
\sqrt{f(r_{b})} = \sqrt{8 G Q^2 \log\left( \frac{r_{G}}{r_{b}} \right)} =   \sqrt{16 G^2 Q^2 \mu^2} = 4 G Q \mu ~ .
\ena
Using the equation for $\beta =1/T$ (\ref{EqSec2beta}), $Q$ can be written as 
\bea
\frac{1}{T} = \sqrt{f(r_{b})} \beta_{0} = 4G Q \mu \times \frac{\pi r_{b}}{2G Q^2} e^{2G \mu^2} =  \frac{2 \pi \mu r_{b}}{ Q} e^{2 G \mu^2} \notag \\
\Leftrightarrow ~~ Q =  2 \pi \mu r_{b} T e^{2 G \mu^2} ~ . \hspace{2cm}
\ena
The free energy $F_{BG}$ can be obtained by the on-shell action (\ref{EqSec2onshell}) and the energy can also be obtained by (\ref{EqSec2BGenergy}). The entropy is obtained by using the relation $F = E-TS-\mu Q$. Therefore, thermodynamic quantities of BG saddles are written as \\
\underline{Thermodynamic  quantity  of  BG  saddle} 
\footnote{
In \cite{Miyashita20242}, the energy was wrongly set to zero and the energy contribution to the free energy was missing. That is, the energy $E^{\rm (wrong)}_{BG}$ and the free energy $F_{\rm BG}^{\rm (wrong)}$ in \cite{Miyashita20242} are related to the true thermodynamic quantities as
\beann
E^{\rm (wrong)}_{BG} = 0 ~ , \hspace{6.05cm} \\
F^{\rm (wrong)}_{BG} = -T S_{BG} -\mu Q_{BG} =- \frac{\pi r_{b} T }{2G} (1 + 4 G \mu^2 ) e^{2G \mu^2} ~ .
\enann
As I wrote in the previous footnote, the linearity of the free energy is not changed by this modification and the conclusions of \cite{Miyashita20242} are unchanged. The transition temperature and the phase diagram are slightly modified.
}
\bea
F_{\rm BG}  =  -\frac{\pi r_{b} T }{2 G} e^{2G \mu^2} ~ , \hspace{2.5cm}\\
E_{\rm BG} = \frac{\sqrt{ f(r_{b})} }{4G} = 2 \pi \mu^2 r_{b} T e^{2 G \mu^2} ~ , \hspace{0.5cm} \\
S_{\rm BG} = \frac{2\pi r_{G}}{4 G} =  \frac{\pi r_{b}}{2 G} e^{2G \mu^2} ~ , \hspace{1.65cm} \label{EqSec3BGent} \\
Q_{\rm BG} =  2 \pi \mu r_{b} T e^{2 G \mu^2} ~ . \hspace{2.55cm}
\ena
Similarly, we can compute those of empty saddles; \\
\underline{Thermodynamic  quantity  of  empty saddle} 
\bea
F_{\rm empty}  =  -\frac{\sqrt{ f(r_{b})} }{4G} = \frac{-1}{4G} ~ , \hspace{0.4cm}\\
E_{\rm empty} = -\frac{\sqrt{ f(r_{b})} }{4G} = \frac{-1}{4G} ~ , \hspace{0.4cm} \\
S_{\rm empty} =0 ~ , \hspace{2.9cm} \\
Q_{\rm empty} =  0 ~ .  \hspace{2.95cm}
\ena
One problem, or property, pointed out in \cite{Miyashita20242} is that the heat capacity of the BG saddle is exactly zero, and it does not seems to be thermodynamically stable. Therefore, I used quotation marks $''\cdots''$ for the thermodynamic terms like ``transition temperature'' or ``phase diagram''. There I suspected that one-loop corrections may make its heat capacity (slightly) positive.
\footnote{
Similar to the empty saddle whose heat capacity is zero at the zero-loop order, but is positive when we include the contribution of fluctuations (or ordinary matter fields).
}
In this paper, I assume this to be true and remove the quotation marks for thermodynamic terms.

The system exhibits Hawking-Page phase transitions just like higher dimensional systems, not between the empty phase and the BH phases,  but between the empty phase and the BG phase. Equating $F_{\rm empty}$ and $F_{\rm BG}$, we can obtain the transition temperature $T_{\rm tr}$;
\bea
F_{\rm empty} = F_{\rm BG}  ~~~ \Rightarrow ~~~ T_{\rm tr} = \frac{1}{2\pi r_{b}} e^{-2G \mu^2}  ~ . \label{EqSec3Ttr}
\ena
The behavior of the free energies and the phase diagram is shown in Figure \ref{FIG2}. 
                                                 %
\iffigure
\begin{figure}[h]
\begin{center}
	\includegraphics[width=7.5cm]{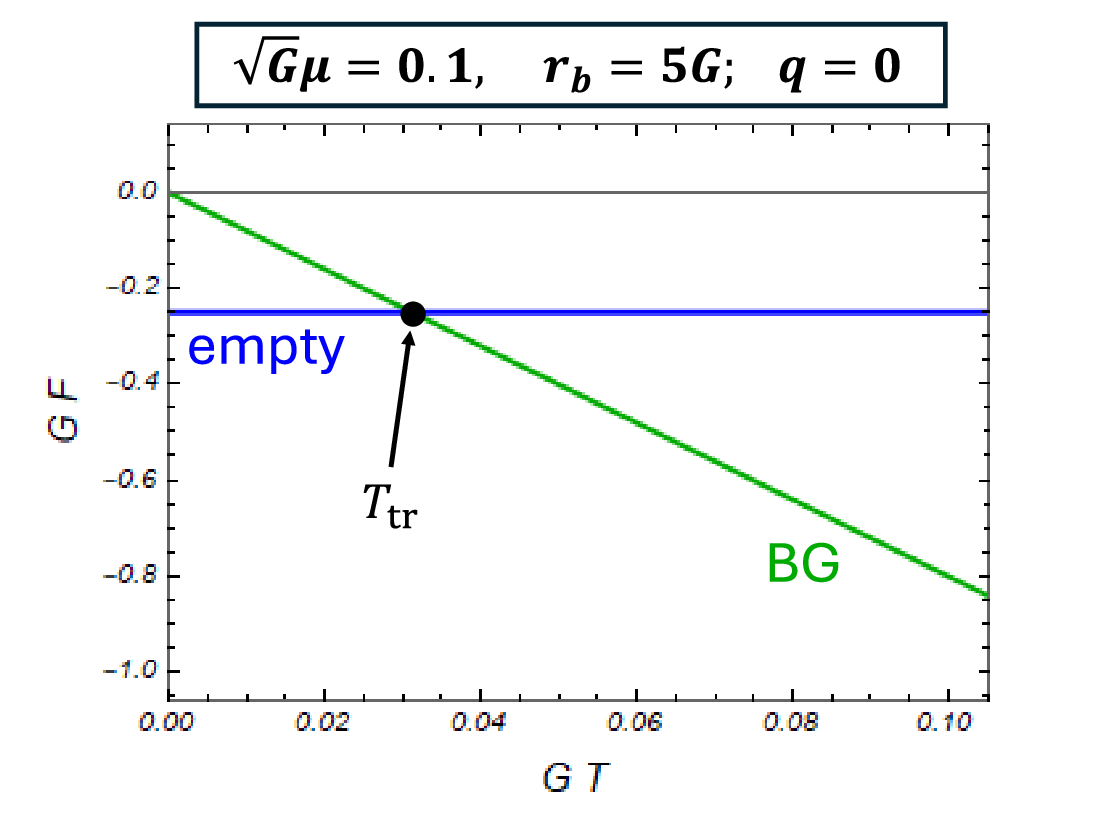} ~~ 	\includegraphics[width=7.5cm]{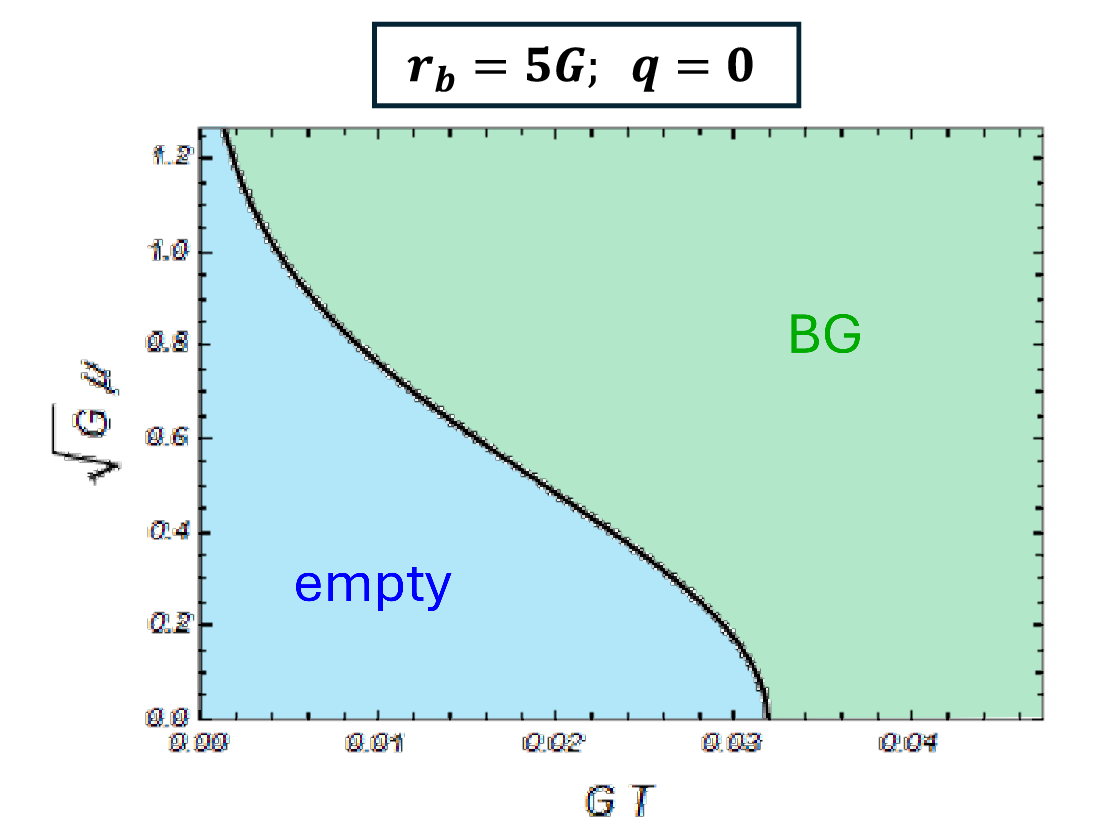} 
	\caption{ \textbf{(left)} Behavior of free energies of  $q=0$. $F_{\rm empty}$ is constant and $F_{\rm BG}$ is a linear function. The point of intersection of the lines corresponds to the transition temperature $T_{\rm tr}$ (\ref{EqSec3Ttr}).  \textbf{(right)}  Phase diagram of $q=0$. The low temperature phase is the empty phase and the high temperature phase is the BG phase. The phase boundary is given by the curve $T = T_{\rm tr}(\mu, r_{b})$ (\ref{EqSec3Ttr}). }
\label{FIG2}
\end{center}
\end{figure}
\fi
                                                 %
As shown in the left panel, for a fixed $\mu$, the low temperature region is dominated by the empty saddle and the high temperature region is dominated by the BG saddle. This property holds for any value of $\mu$. The resulting phase diagram is shown in the right panel. Thus, the phase structure of the EMS system with $q=0$ (or the EM system) in three dimensions with $\Lambda=0$ is similar to that of the EM system in higher dimensions,
\footnote{
Both for the case of AdS boundary \cite{CEJM1999} and for the case of Dirichlet boundary \cite{Miyashita20241}.
}
although there are no BHs. The high temperature phase is now replaced by the BG phase. 

\subsection{Dissimilarities}
In the previous subsection, I emphasized the similarities to the thermodynamic properties of the EM system in the well-known settings, i.e. with the AdS boundary condition. They are the facts that the Hawking-Page phase transition is the same as in the $d$-dimensional $(d\geq3)$ AdS case, the behavior of the free energies is similar to that in the three-dimensional AdS case, and the phase diagram is also similar to the AdS case. However, there are also many dissimilarities that are worth highlighting in order to clearly see the difference between the $q=0$ (or EM) case and the finite $q$ case. One of them is, of course, that the system does not admit BH saddles, but admits BG saddles as explained previously. I list the other dissimilarities below, some of which I did not emphasize in the previous subsection or in \cite{Miyashita20242};
\begin{itemize}
\item BG entropy $S_{\rm BG}$ does not depend on temperature $T$. \\
In standard gravitational thermodynamics, as we increase $T$, the entropy or horizon area increases. However,  in this case the entropy is given by $S_{\rm BG} = \frac{\pi r_{b}}{2 G} e^{2G \mu^2} $ (\ref{EqSec3BGent}). For a fixed chemical potential $\mu$, no matter how we change $T$ (above the Hawking-Page point), the horizon area does not change! (Note that this does {\it not} mean that the geometry does not change.)

\item BG free energy $F_{BG}$ is linear to $T$. The heat capacity is zero. \\
Related to the previous fact, the BG free energy $F_{BG}$ is a linear function of $T$, and thus the heat capacity is exactly zero at the zero-loop order. This is a peculiar property since usually the dominant BH (or BG) saddles have positive specific heat even at the zero-loop and give us well-defined thermodynamic descriptions. Although I suspect this {\it problem} can be solved by one-loop corrections, it is still a somewhat remarkable {\it property} of this saddle or system.

\item No extremal horizons. The empty phase is always dominant at low temperatures. \\
One remarkable property of charged systems is that they generally have extremal BH (or BG) saddles. Although the existence of the saddles does not necessarily imply dominance in the GPI, their phase appears when the chemical potential $\mu$ is sufficiently high in EM system with AdS boundary condition. 
\footnote{
The case of Dirichlet boundary condition in higher dimensions is subtle. When $\Lambda < 0$, the extremal horizon states appear the same as in  the AdS boundary case. When $\Lambda = 0$, they do not appear.
}
As shown in Figure \ref{FIG2} \textbf{(right)}, however, we do not have an extremal BG phase in this system. The zero and low temperature phase is always the empty phase. 
\end{itemize}

\section{Thermodynamics of Finite $q$}
Let's consider the case of finite $q$. Bearing in mind that the phase structure of the $q=0$ case (or the EM case) is similar to that of the EM system in higher dimensions with AdS boundary, turning on the charged scalar field may correspond to considering ``holographic'' superconductors \cite{Gubser2008, HHH20081, HHH20082} . Therefore, it is natural to expect the system to have a similar phase structure to the holographic superconductors. In this section, I confirm that this is true by showing that there exist many types of saddles or phases in this system.
\subsection{saddles}
When the coupling $q$ is turned on, new types of saddles appear; the boson star saddle, the boson star-PL saddle and the hairy BG saddle. The first one was found and studied in \cite{KSB2020}. The newly constructed saddles in this paper are the latter two. So there are five types of saddles in total; \\
 $\cdot$ empty saddle ~ (already explained in Section 3) \\
 $\cdot$ BG saddle ~ (already explained in Section 3) \\
 $\cdot$ boson star saddle ~ (to be explained in 4.1.1.) \\
 $\cdot$ boson star-PL saddle ~ (to be explained in 4.1.2.) \\
 $\cdot$ hairy BG saddle ~ (to be explained in 4.1.3.) \\
The last three must be constructed numerically. I will explain their construction and properties separately below.

\subsubsection{boson star saddle}
In this case, the coordinate range for $r$ is $(0, r_{b})$, the same as for the empty saddle. However, there is a condensate of the charged scalar field around the center. Let's expand the variables $\Delta, f, a, \varphi$ around the center as
\bea
\Delta(r) = \sum_{k=0} \Delta_{k} r^k, ~~~~~ f(r) = \sum_{k=0} f_{k} r^k, ~~~~~ a(r) = \sum_{k=0} a_{k} r^k, ~~~~~ \varphi(r) = \sum_{k=0} \varphi_{k} r^k ~ .
\ena
First, by imposing the regularity condition $\Delta'(0)=0, f'(0)=0, a'(0)=0, \varphi'(0)=0,$ we obtain $\Delta_{1} = f_{1} = a_{1}= \varphi_{1} =0$. Then, substituting the above expansion into the EOM (\ref{EqSec2EOMdelta}) - (\ref{EqSec2EOMphi}), we obtain
\bea
\Delta(r) = \Delta_{0} + \frac{ 16 \pi G q^2 a_{0}^2 \varphi_{0}^2 }{f_{0}^2} r^2 + \cdots ~ , \\
f(r) = f_{0} - \frac{ 8\pi G q^2 a_{0}^2 \varphi_{0}^2 }{ \Delta_{0} f_{0} } r^2 + \cdots ~ , \hspace{0.25cm} \\
a(r) = a_{0} + \frac{2 \pi q^2 a_{0} \varphi_{0}^2 }{f_{0}} r^2 + \cdots ~ , \hspace{0.5cm} \\
\varphi(r) = \varphi_{0} - \frac{ q^2 a_{0}^2 \varphi_{0} }{4 \Delta_{0} f_{0}^2} r^2 + \cdots ~ ,  \hspace{0.9cm}
\ena
where the subleading terms can also be determined order by order. This means that the behavior of the fields is completely determined by the four constants $(\Delta_{0}, f_{0}, a_{0}, \varphi_{0})$. However, the first two are actually redundant; \\
$\cdot$ $\Delta_{0}$ corresponds to the freedom of time rescaling. If we rescale $\Delta \to \alpha^2 \Delta, a \to \alpha a, t\to t/\alpha$, the metric (\ref{EqSec2metric}) and the gauge field (\ref{EqSec2gauge}), and the EOM (\ref{EqSec2EOMdelta}) - (\ref{EqSec2EOMphi}) are unchanged. Therefore, we can set $\Delta_{0} =1$. \\
$\cdot$ $f_{0}$ corresponds to the freedom of a certain simultaneous rescaling of  coordinates. If we rescale $r \to \alpha r, t \to t/\alpha, \phi \to \phi/\alpha$, and $f \to \alpha^2 f, a \to \alpha a$, the metric, gauge field, and the EOM are unchanged. So we can set $f_{0} =1$.
\footnote{
This freedom of rescaling appears only in three dimensions, not in higher dimensions. In \cite{KSB2020}, $f_{0}$ (which is what they call $g_{0}$ in their paper) was considered as a new degree of freedom specific to three dimensions. However, I believe they missed this rescaling symmetry and $f_{0}$ just represents this redundancy.
}
\\
Therefore, there are only two parameters $(a_{0}, \varphi_{0})$. However, we can see that generic choices of parameters do not lead to solutions satisfying the boundary condition $\varphi(r_{b})=0$ (\ref{EqSec2bdyphi}). We need to shoot appropriate points on the parameter space $(a_{0}, \varphi_{0})$ so that the resulting numerical solution satisfies the boundary condition. The appropriate points form a curve on the space $(a_{0}, \varphi_{0})$. An example of  such a curve on the space and the numerical solution are shown in Figure \ref{FIG3}.
\footnote{
Note that there are also solutions that satisfy the boundary condition $\varphi(r_{b})=0$ (\ref{EqSec2bdyphi}) but have multiple nodes in the scalar configuration. The figure \ref{FIG3} is for the 0 node case. (For 1 node solutions, see \cite{KSB2020}.) In general, hairy solutions, such as hairy soliton or hairy BH, can have multiple node solutions. In fact, I have found multiple node solutions not only for this boson star saddle, but also for boson star-PL saddles and hairy BG saddles. However, as a rule of thumb for systems with hairy solutions, only the lowest, mostly zero, node solutions are stable in any sense, i.e. they are perturbatively stable, or thermodynamically stable, and so on. Therefore, in this paper I will concentrate only on the 0 node solutions. \label{foot1}
}
                                                 %
\iffigure
\begin{figure}[h]
\begin{center}
	\includegraphics[width=7.5cm]{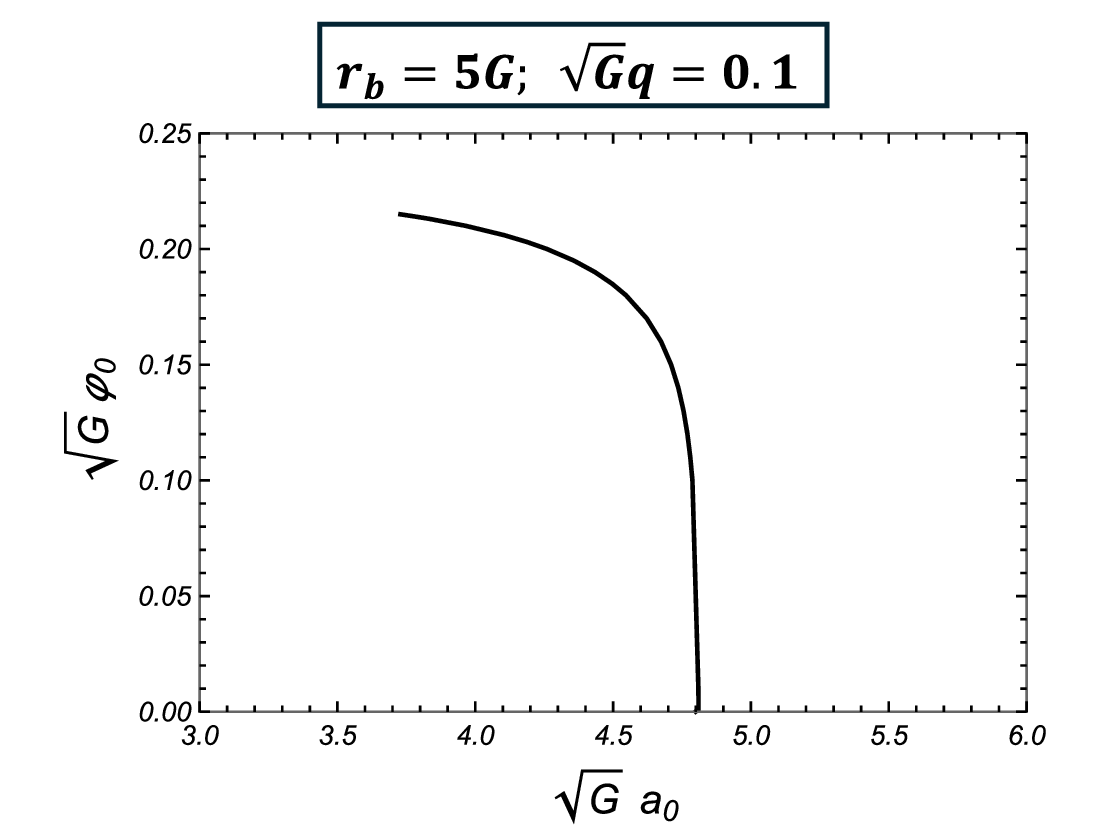} ~~ 	\includegraphics[width=7.5cm]{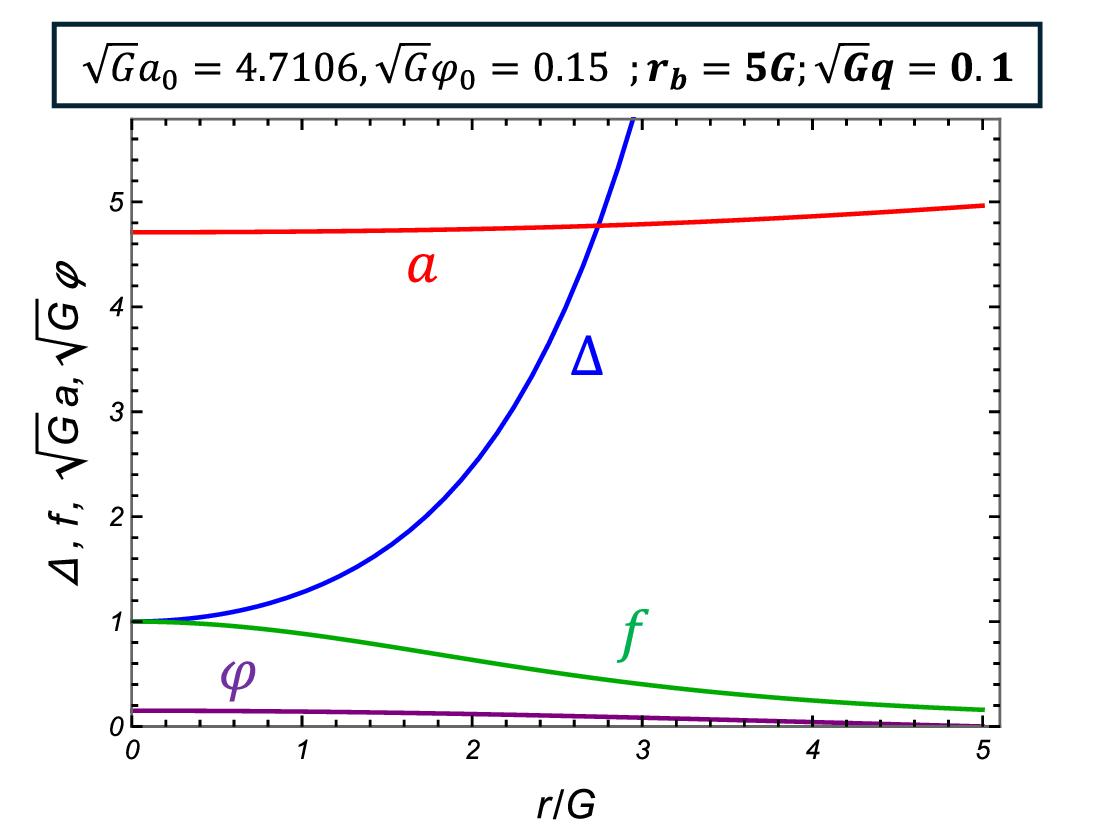} 
	\caption{ \textbf{(left)} The solution curve in the space of $(a_{0}, \varphi_{0})$ for $r_{b}=5 G$ and $\sqrt{G} q=0.1$. Solutions corresponding to the other points neither satisfy the boundary condition $\varphi(r_{b})=0$ nor have $\varphi(r)$ more than one zero. (For the latter, see the footnote \ref{foot1}.) The endpoint of the curve corresponds to the solution with $f(r_{b}) \to 0$. Therefore, numerically determining the exact location of the endpoint is a somewhat difficult task. ~~ \textbf{(right)}  Example of a numerical solution. (I set $\sqrt{G} a_{0} = 4.7106, \sqrt{G} \varphi_{0} = 0.15, r_{b}=5 G, \sqrt{G}q=0.1$.) }
\label{FIG3}
\end{center}
\end{figure}
\fi
                                                 %
The right panel shows an example of a numerical solution. $a(r)$ and $\Delta$ are monotonically increasing, while $\varphi(r)$ and $f(r)$ are monotonically decreasing. This property holds for other initial conditions. A solution curve in the space of $(a_{0}, \varphi_{0})$ is shown in the left panel. $\varphi_{0} = 0$ corresponds to the empty saddle (\ref{EqSec3empty}). As explained in \cite{KSB2020}, the curve has an endpoint since $f(r_{b})$ goes to zero as we increase $\varphi_{0}$ and it will be zero at some finite $\varphi_{0}$. This is the endpoint of the boson star saddles. However, this is not an endpoint of solutions. This is where a new type of solution appears, which I will call ``boson star-PL''saddle.

\subsubsection{boson star-PL saddle (and boson star saddle)}
As $\varphi_{0}$ is increased, $f(r_{b})$ goes to zero. One might think that this corresponds to the formation of a horizon or some kind of singularity at or around $r=r_{b}$. This is not the case in this system. Let's call the point where $f(r) = 0$ as $r = r_{max}$. This can be any positive value, depending on values of the shooting parameters. If we carefully analyze the behaviors of the variables around $r=r_{max}$, we can see that they behave as
\bea
\Delta(r)  = \frac{\Delta_{-1} }{ ( r_{max} - r ) } + \cdots ~ , \hspace{7.8cm} \label{EqSec4rmaxdelta1} \\
f(r) =   f_{1} ( r_{max} - r )  + \cdots ~ ,  \hspace{7.55cm}   \\
a(r) = a_{max} - \left( \sqrt{\frac{f_{1} \Delta_{-1} }{ G r_{max}}} \right) ( r_{max} - r )^\frac{1}{2} + \cdots ~ , \hspace{4.15cm} \label{EqAExpndPL} \\
\varphi(r) =  \varphi_{\max} + \left( \sqrt{ \frac{  f_{1}^2 \Delta_{-1} - 32 \pi G q^2 r_{max} a_{max}^2  \varphi_{max}^2 }{8 \pi G r_{max} f_{1}^2 \Delta_{-1}  } }  \right) ( r_{max} - r )^\frac{1}{2}  + \cdots ~ .  \label{EqPhiExpndPL}
\ena
                                                 %
\iffigure
\begin{figure}[t]
\begin{center}
	\includegraphics[width=7.5cm]{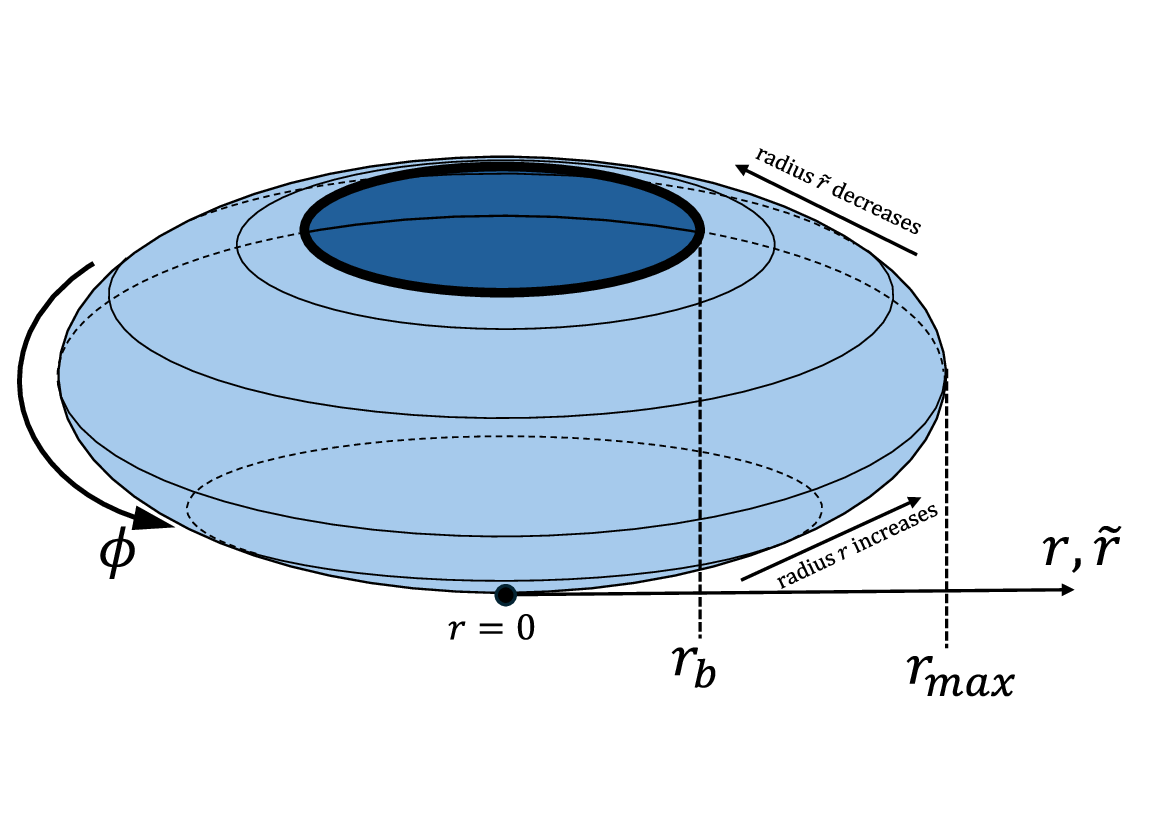} ~~ 	\includegraphics[width=7.5cm]{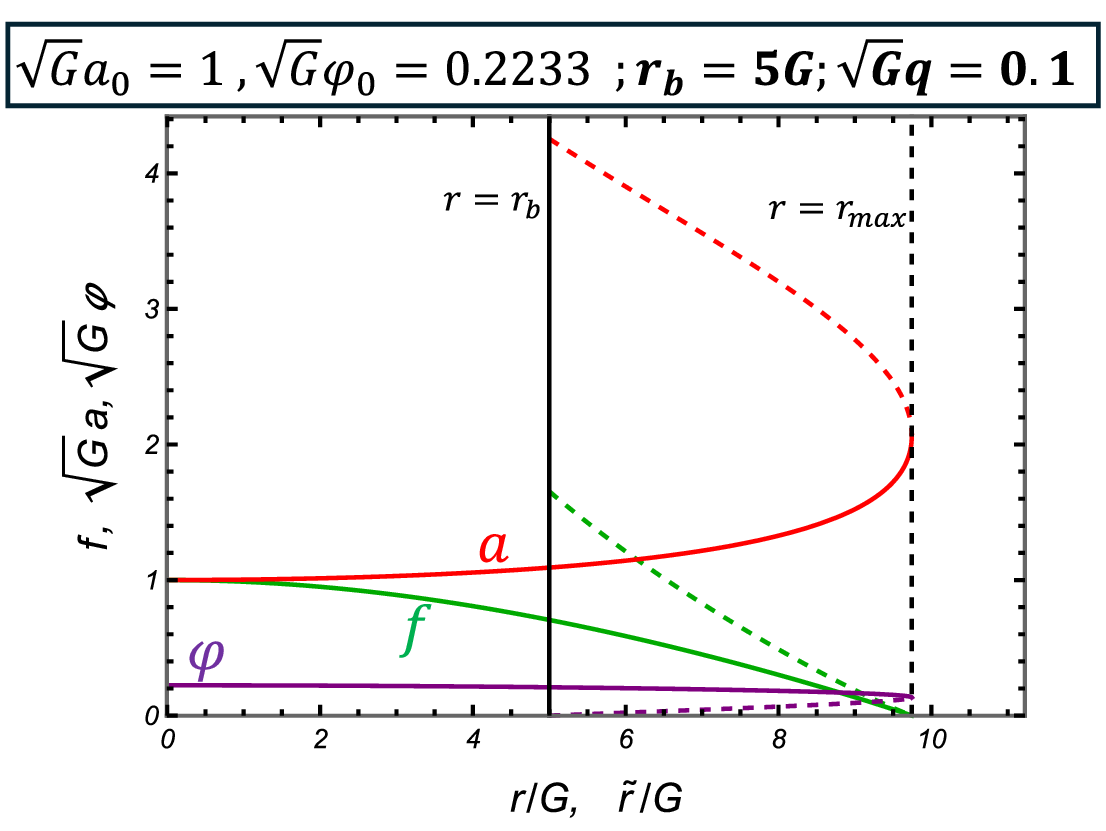} 
	\caption{ \textbf{(left)} Schematic picture of the $r(\tilde{r})-\phi$ section of the boson star-PL saddle. It is covered by two coordinate systems; the lower hemisphere is covered by $(t, r, \phi)$ and the upper hemisphere is covered by $(t, \ti{r}, \phi)$. The upper hemisphere has a hole whose boundary is the boundary $\ti{r}=r_{b}$. \textbf{(right)} Example of a boson star-PL solution. For pictorial clarity, $\Delta$ is not plotted. The solid curves represent the behaviors in the lower hemisphere thus they are functions of $r$. The dashed curves represent those in the upper hemisphere thus they are functions of $\ti{r}$. As is written in eq. (\ref{EqSec4rmaxdelta1}) and (\ref{EqSec4rmaxdelta2}), $\Delta$ diverges at $r=r_{max}$ in both the lower hemisphre and the upper hemisphere. (I set $\sqrt{G} a_{0} = 1, \sqrt{G} \varphi_{0} = 0.2233, r_{b}=5 G, \sqrt{G}q=0.1$.) }
\label{FIG4}
\end{center}
\end{figure}
\fi
                                                 %
The important points are that (i) $g_{tt}(r) = \Delta(r) f(r) $ is not singular, and (ii) although $g_{rr} \to \infty$, this singularity can be removed by appropriate coordinate transformations, e.g. $ - \sqrt{r_{max} - r} = \rho $, which leads to $g_{\rho\rho} \simeq \frac{4}{f_{1}}$ around $r = r_{max}$. So  $r_{max}$ is neither a horizon nor a singularity and we can extend this geometry beyond $r= r_{max}$ by considering the $\rho>0$ region. (Recall that I defined $\rho = -\sqrt{r_{max} - r}$ and the $\rho <0 $ region corresponds to the $r<r_{max}$ region.) If we define a new coordinate $\ti{r}$ by $\rho = \sqrt{r_{max} - \ti{r}}$, it becomes a new radial coordinate but covers a different region of the geometry. In this coordinate system, the expansion of the variables can be written as
\footnote{
Here, the signs of the second terms in the expansions of $a$ and $\varphi$ differ from those in equations (\ref{EqAExpndPL}) and (\ref{EqPhiExpndPL}). This discrepancy arises because the expansions in (\ref{EqAExpndPL}) and (\ref{EqPhiExpndPL}) are valid for $\rho \in \mathbb{R}$ under the coordinate transformation $\rho = -\sqrt{r_{\text{max}} - r}$. When we instead use the coordinate transformation $\rho = \sqrt{r_{\text{max}} - \tilde{r}}$ with $\rho > 0$, an additional minus sign appears in front of the second term due to the change in the sign of $\rho$.
}
\bea
\Delta(\ti{r})  = \frac{\Delta_{-1} }{ ( r_{max} - \ti{r} ) } + \cdots ~,  \hspace{7.8cm} \label{EqSec4rmaxdelta2}\\
f (\ti{r}) =   f_{1} ( r_{max} - \ti{r} )  + \cdots ~ ,  \hspace{7.55cm}  \\
a (\ti{r}) = a_{max} + \left( \sqrt{\frac{f_{1} \Delta_{-1} }{ G r_{max}}} \right) ( r_{max} - \ti{r} )^\frac{1}{2} + \cdots ~ , \hspace{4.15cm} \\
\varphi (\ti{r}) =  \varphi_{\max} - \left( \sqrt{ \frac{  f_{1}^2 \Delta_{-1} - 32 \pi G q^2 r_{max} a_{max}^2  \varphi_{max}^2 }{8 \pi G r_{max} f_{1}^2 \Delta_{-1}  } }  \right) ( r_{max} - \ti{r} )^\frac{1}{2}  + \cdots  ~ . 
\ena
Using this property, we can construct a new type of saddle, which I call the ``boson star-PL'' saddle;
\footnote{
PL stands for Python's Lunch \cite{BGPS2020}. Here, I have nothing to say about it except that the shape of the saddles is similar to the Python's Lunch. 
}
These are solutions with a regular center, with a ``bulge'' $r_{max} > r_{b}$, and satisfy the boundary condition $\varphi(r_{b})=0$ (\ref{EqSec2bdyphi}) after $\ti{r}$ returns from $r_{max}$ to $r_{b}$. A schematic picture of the $r(\ti{r})-\phi$ section of this saddle and an example of solution are shown in Figure \ref{FIG4}.
Although the topology of the $r(\ti{r}) -\phi$ section is still a disc, its geometry looks more like a sphere with a hole as shown in the left panel. The right panel shows the behavior of the variables in the lower and upper hemispheres, represented by solid and dashed curves, respectively.
Again, boson star-PL solutions form a curve on the space $(a_{0}, \varphi_{0})$, and in fact it is smoothly connected to the curve of boson star solutions. An example of the solution curve of the boson star-PL, together with that of the boson star, is shown in Figure \ref{FIG5}. 
                                                 %
\iffigure
\begin{figure}[t]
\begin{center}
	\includegraphics[width=7.5cm]{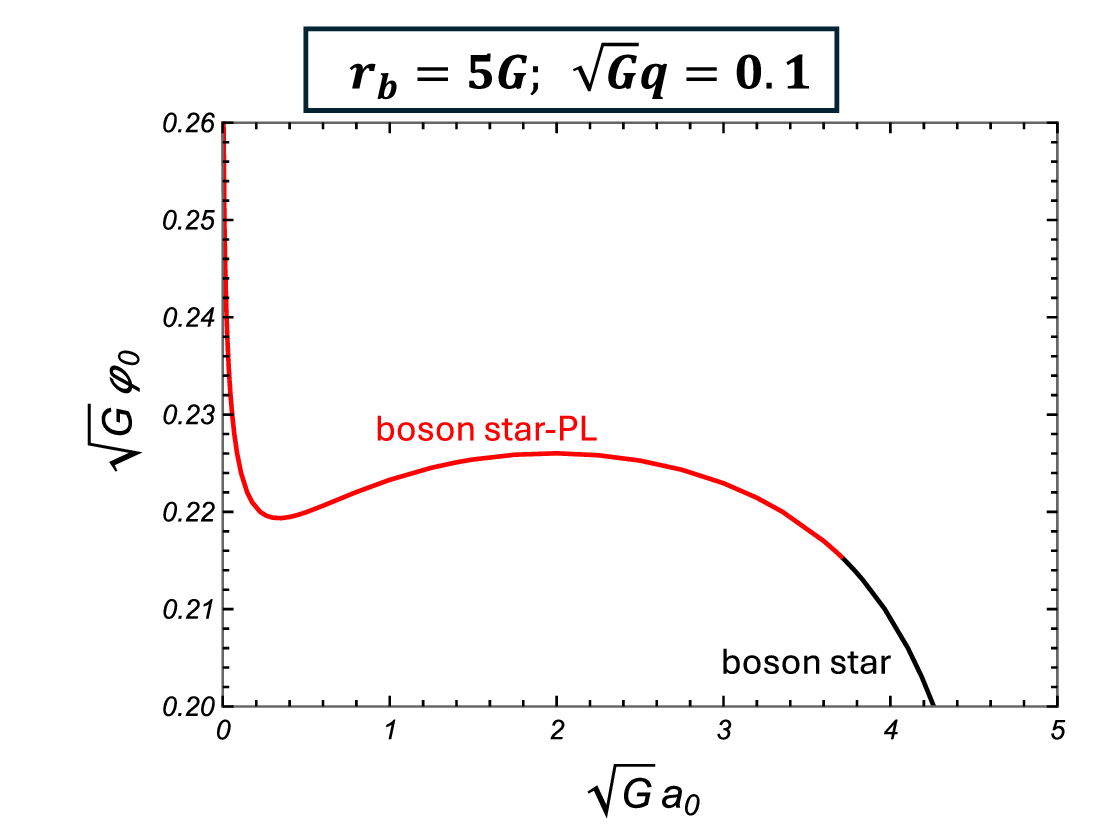} ~~ 	\includegraphics[width=7.5cm]{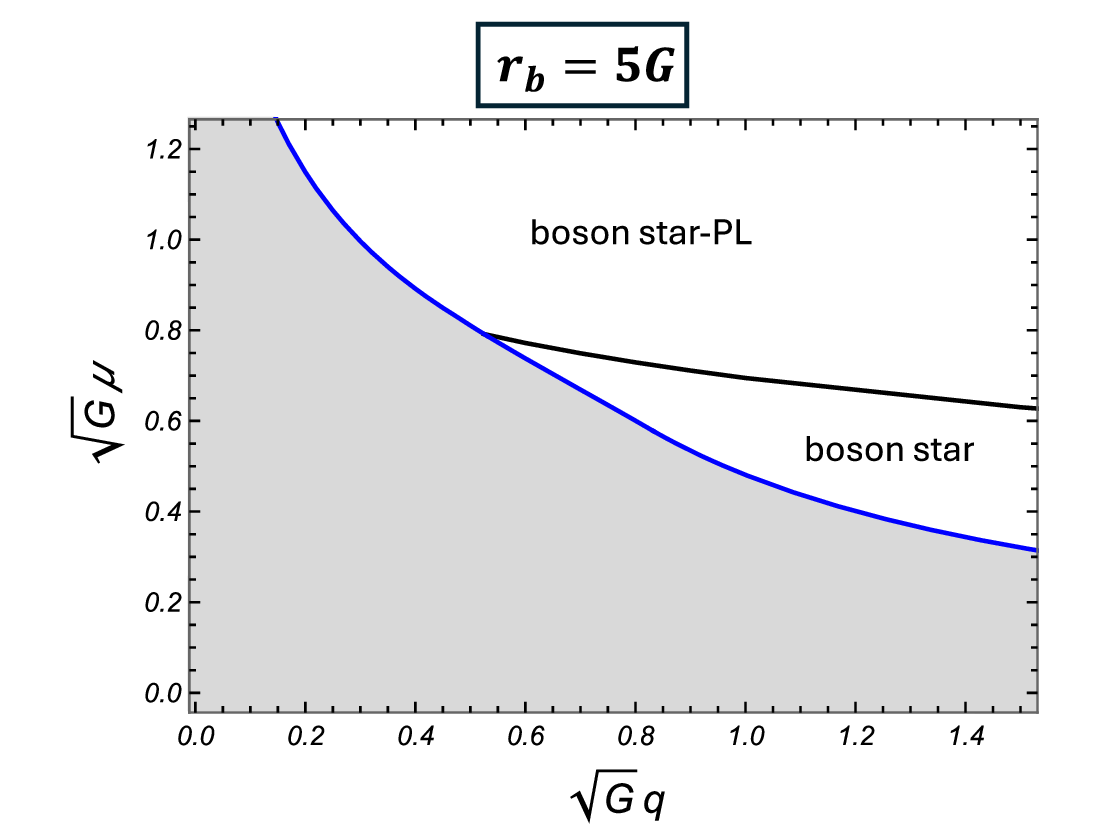} 
	\caption{ \textbf{(left)}  The solution curve for boson star (black) and boson star-PL (red) in the space of $(a_{0}, \varphi_{0})$ for $r_{b}=5 G$ and $\sqrt{G} q=0.1$. Solutions corresponding to the other points neither satisfy the boundary condition $\varphi(r_{b})=0$ nor have $\varphi(r)$ more than one zero. The endpoint of the boson star curve is smoothly connected to the boson star-PL curve. The boson star-PL curve goes to $\varphi_{0} \to \infty$ as it approaches $a_{0} =0$.  \textbf{(right)} The region where the boson star free energy or the boson star-PL free energy is lower than that of empty saddle on the space of $(q, \mu)$. The point where two curves intersect is $(\sqrt{G}q, \sqrt{G}\mu) \simeq (0.525, 0.792)$. }
\label{FIG5}
\end{center}
\end{figure}
\fi
                                                 %
Before discussing thermodynamics in subsection 4.2, it may be good to compare the free energy of empty saddles and that of boson star(-PL) saddles here, since they do not depend on temperature and are easy to compare. It is also shown in Figure \ref{FIG5}. The boson star could be a stable thermal state only when $q \gtrsim 2.625 \frac{\sqrt{G}}{r_{b}}$.
\footnote{
Although the definition of the thermodynamic variable and the boundary size are different, Figure \ref{FIG5} \textbf{(right)} (or Figure \ref{FIG6}) corresponds to the figure 7 of \cite{KSB2020}. In particular, the boson star region of Figure \ref{FIG5} \textbf{(right)} corresponds to the green region of their figure. According to their figure, it seems that the lowest value of $q$ at which the boson star could be stable seems to be around $q \simeq 3.5 \frac{\sqrt{G}}{r_{b}} $, which is different from mine. I found that we should be very careful with the numerical analysis around this intersection point, otherwise an order one error could occur. I believe this is the source of the difference. (See also the caption of Figure \ref{FIG6}.)
}
For relatively small value of the coupling constant, there is no way for the boson star geometry to appear. Instead, only the boson star-PL geometry could appear. 
                                                 %
\iffigure
\begin{figure}[t]
\begin{center}
	\includegraphics[width=5.3cm]{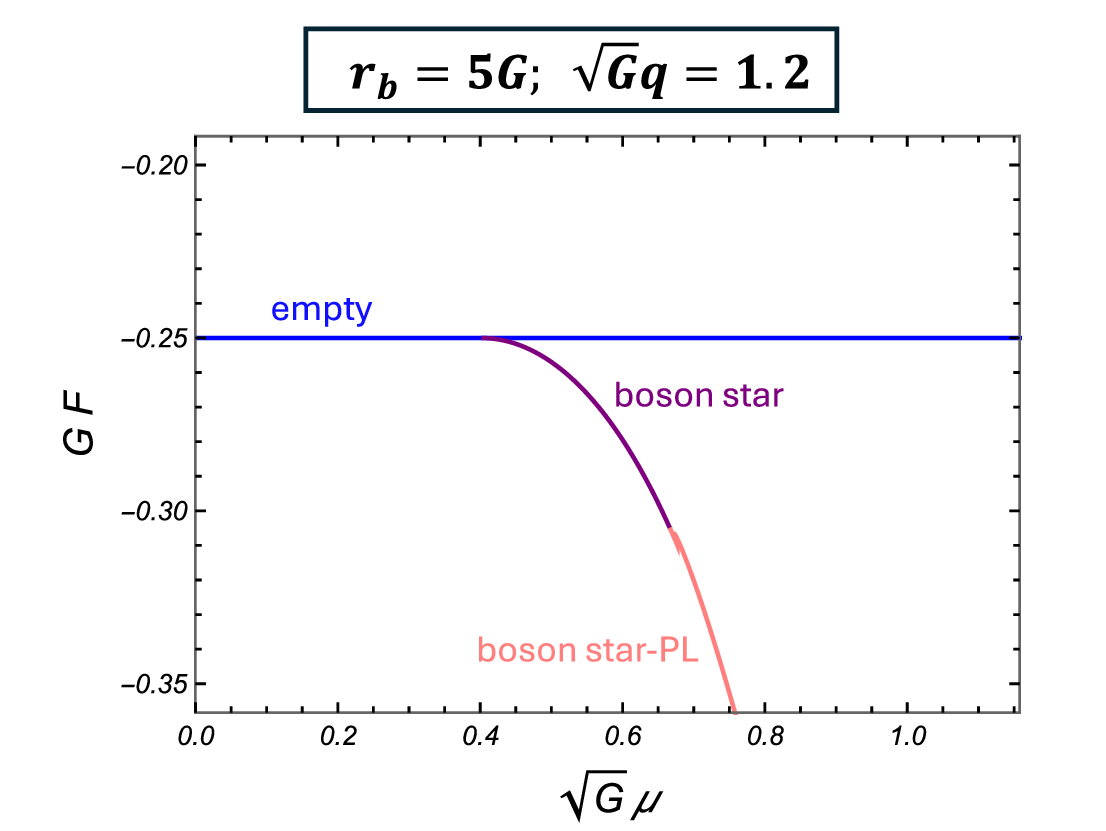} ~ 	\includegraphics[width=5.3cm]{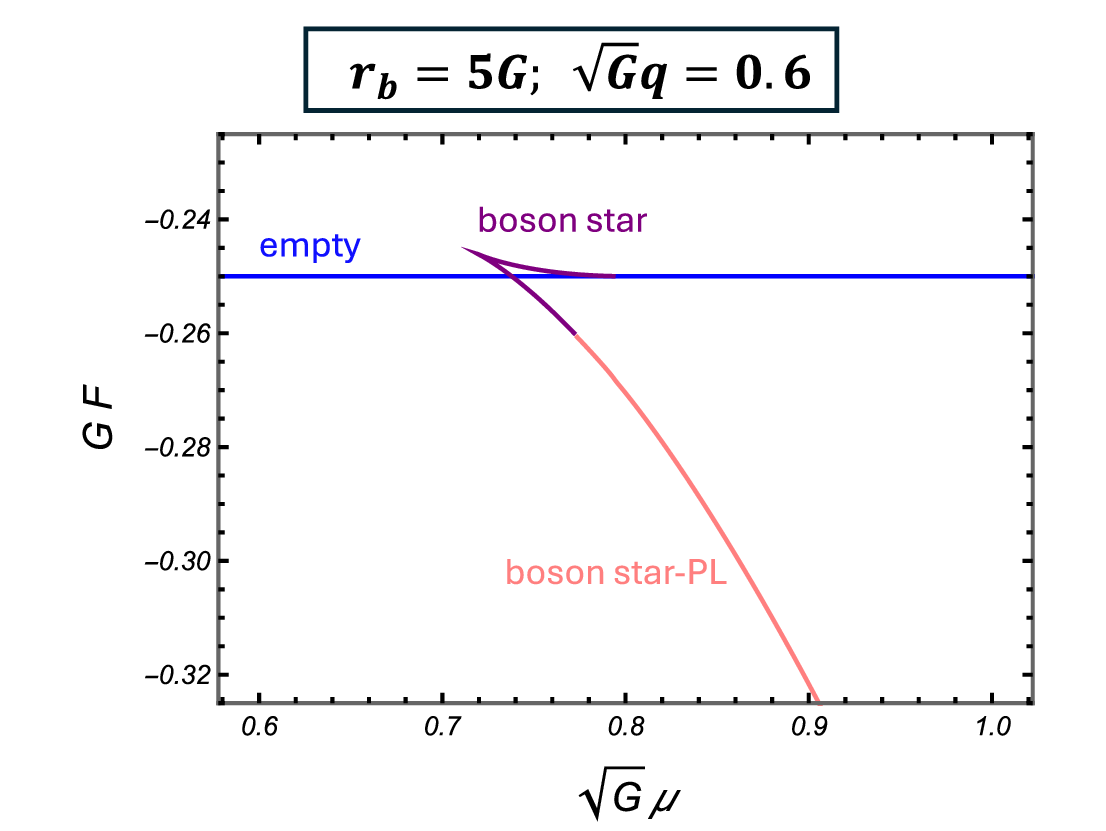}  ~ 	\includegraphics[width=5.3cm]{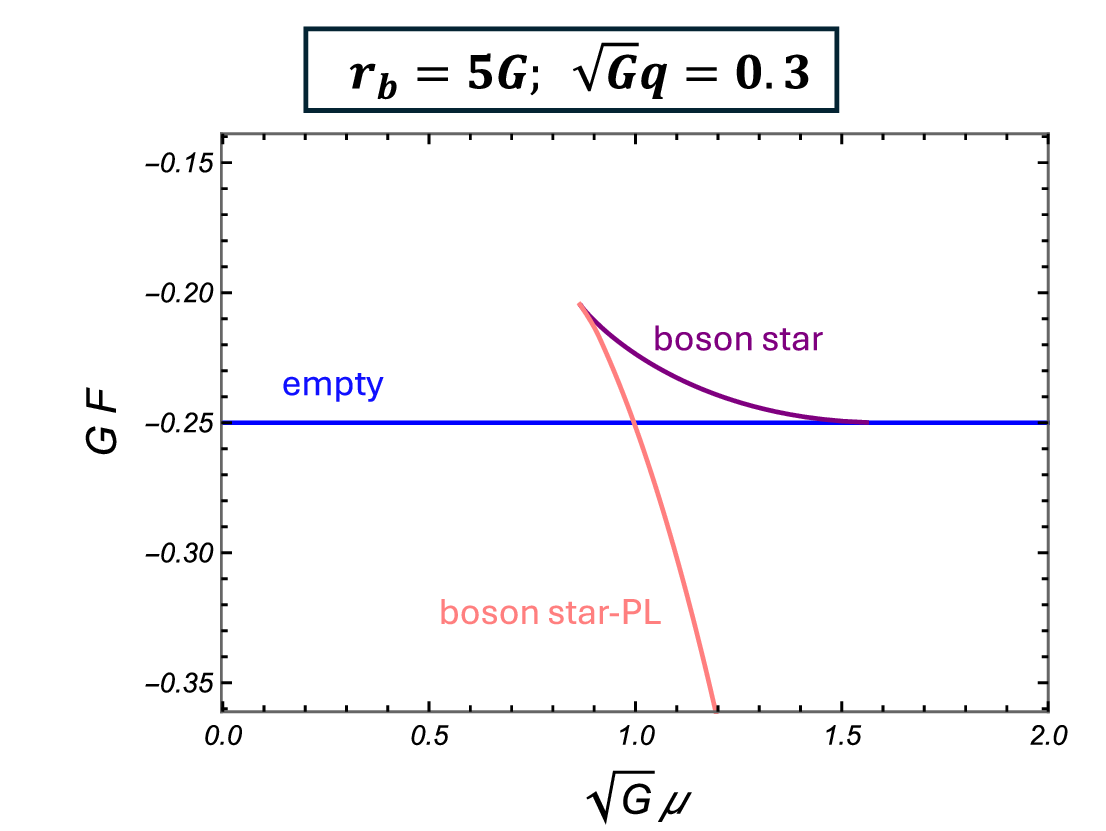} 
	\caption{ Behavior of free energy of boson star and boson star-PL saddles. \textbf{(left)} is that of relatively high $q$,  \textbf{(middle)} is that of intermediate value of $q$, and \textbf{(right)} is that of relatively low $q$. When $q$ is high enough, the system exhibits the second order phase transition between empty saddle and boson star saddle as \textbf{(left)}. When $q$ is low, it is the first order phase transition as \textbf{(middle)} or \textbf{(right)}. The behavior as \textbf{(right)} appears when $q \lesssim 2.625 \frac{\sqrt{G}}{r_{b}} $. Determining the boundary value of $q$ between \textbf{(left)} and \textbf{(middle)} is numerically difficult. It is probably around $q \sim 4.0 \frac{\sqrt{G}}{r_{b}} $. (I set $\sqrt{G}q=1.2$ for  \textbf{(right)}, $\sqrt{G}q=0.6$ for  \textbf{(middle)}, and $\sqrt{G}q=0.3$ for  \textbf{(right)}. $r_{b}=5 G$ for all three figures. ) \\
Note that there is a small zigzag shape in the boson star-PL branch near the junction in \textbf{(left)}. This kind of behavior may not be acceptable from a thermodynamic point of view. I believe this is an artifact of a lack of numerical accuracy. }
\label{FIG65}
\end{center}
\end{figure}
\fi
                                                 %
\clearpage
Examples of free energy versus the chemical potential are shown in Figure \ref{FIG65}. The order of the phase transition between the empty saddle and the boson star(-PL) saddle depends on $q$. When $q$ is high enough, it is the second order phase transition as Figure \ref{FIG65} \textbf{(left)}. When $q$ is low, it is the first order phase transition as Figure \ref{FIG65} \textbf{(middle)} and \textbf{(right)}. 
In addition to the possible stable saddles, there are other unstable saddles. Although they are not directly related to the following thermodynamic analysis, I explain them in Figure \ref{FIG6} and Table \ref{TAB1}.
                                                 %
\iffigure
\begin{figure}[h]
\begin{center}
	\includegraphics[width=8cm]{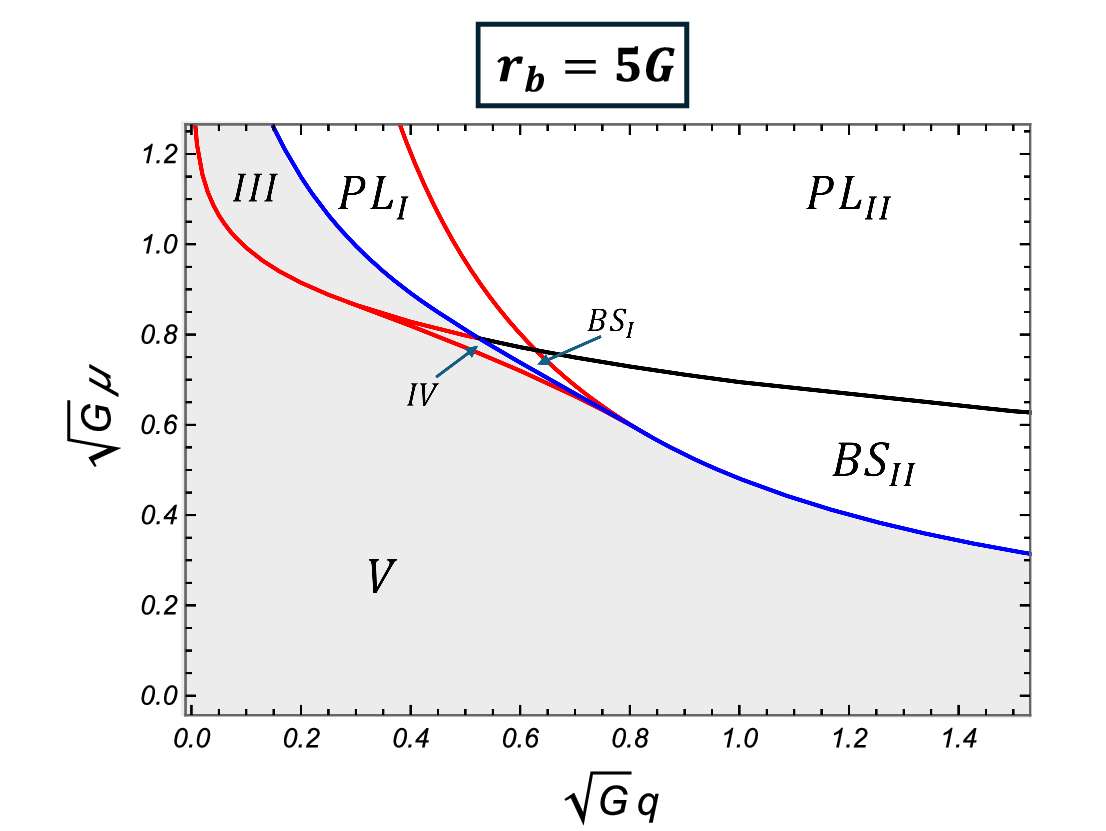} 
	\caption{ The regions classified by the number of boson star and boson star-PL saddles and by the value of their free energy on the space of $(q, \mu)$. The explanation of each region is given in the table \ref{TAB1}. Careful analysis may be required to confirm the existence of the regions $BS_{I}$ and $IV$.   }
\label{FIG6}
\end{center}
\end{figure}
\fi
                                                 %

\begin{table}[h]                
\begin{center}
\begin{tabular}{c|c|c|c}
& \textbf{Number of boson star}	& \textbf{Number of boson star-PL} 	& \textbf{Stable saddle}  \\
\hline \hline
$PL_{I}$		& 1		& 1	& boson star-PL \\
\hline
$PL_{II}$		& 0		& 1	& boson star-PL \\
\hline
$BS_{I}$		& 2     & 0	& boson star\\
\hline
$BS_{II}$		& 1		& 0	& boson star\\
\hline
$III$		      & 1		& 1	& None \\
\hline
$IV$       	& 2		& 0	& None \\
\hline
$V$		       &  0	 & 0	& None \\
\end{tabular}
\caption{Classification in Figure \ref{FIG6}. They are classified by the number of boson star and boson star-PL saddles and by which saddle has lower free energy than that of the empty saddle. Here, ``stable saddle'' means only the one whose free energy is lower free energy than that of the empty saddle. The other saddles may have local thermodynamic stability. And ``stable'' does not necessarily mean global thermodynamic stability. }
\label{TAB1}
\end{center}
\end{table}                

\subsubsection{hairy BG saddle}
Finally, let's consider hairy BG saddles. Suppose the existence of the horizon where $f(r_{G})$ =0, where $r_{G}$ is the position of the horizon. By imposing the regularity at the horizon for other fields, $\Delta(r_{G})= {\rm finite}$,  $\varphi(r_{G})= {\rm finite}$, and $a(r_{G})=0$, the expansion of the variables around $r=r_{G}$ is 
\bea
\Delta(r) = \Delta_{G0} + \frac{8 \pi q^2 \Delta_{G0}^2 \varphi_{G0}^2}{ G a_{G1}^2 r_{G} } (r-r_{G}) + \frac{4\pi q^2 \Delta_{G0}^2 \varphi_{G0}^2 (G a_{G1}^2 +12 \pi q^2 \Delta_{G0} \varphi_{G0}^2) }{ G^2 a_{G1}^4 r_{G}^2} (r-r_{G})^2 + \cdots ~ ,\\
f(r) = - \frac{2 G r_{G} a_{G1}^2}{ \Delta_{G0} } (r-r_{G}) + \frac{ G a_{G1}^2 + 12 \pi q^2 \Delta_{G0} \varphi_{G0}^2 }{ \Delta_{G0} } (r-r_{G})^2 + \cdots ~ , \hspace{4.0cm} \\
a(r) = a_{G1} (r-r_{G}) - \frac{a_{G1}}{2 r_{G}} (r-r_{G})^2 + \cdots ~ , \hspace{8.4cm} 
\\
\varphi(r) = \varphi_{G0} - \frac{q^2 \Delta_{G0} \varphi_{G0} }{ 16 G^2 a_{G1}^2 r_{G}^2 } (r-r_{G})^2 + \cdots ~ . \hspace{8.5cm} 
\ena
Note that this expression also holds for BH horizons. However, in that case, we see that $f' = -\frac{2 G r_{G} a_{G1}^2}{\Delta_{G0}}$ is always negative for any value of the horizon radius. Therefore, there are no BH horizons \cite{KSB2020, Ida2000}. But BG horizons can exist.

\begin{figure}[t]
\begin{center}
	\includegraphics[width=7.cm]{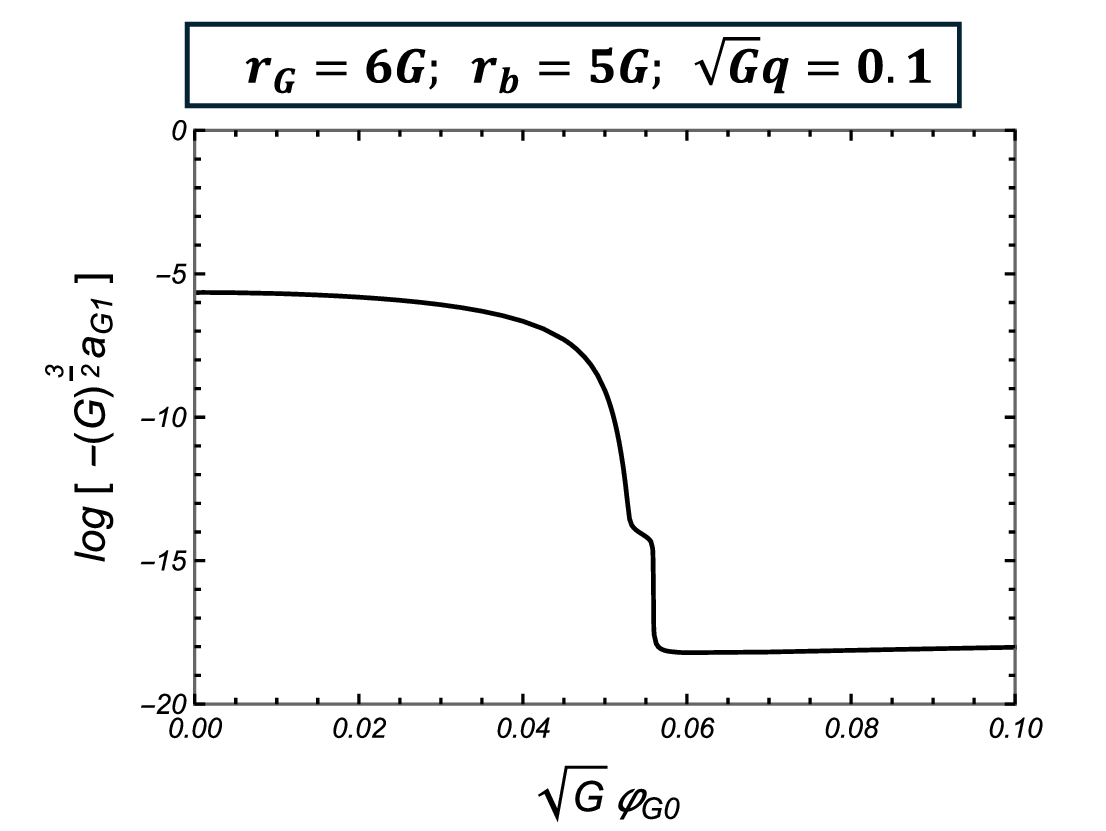} ~~ 	\includegraphics[width=7.cm]{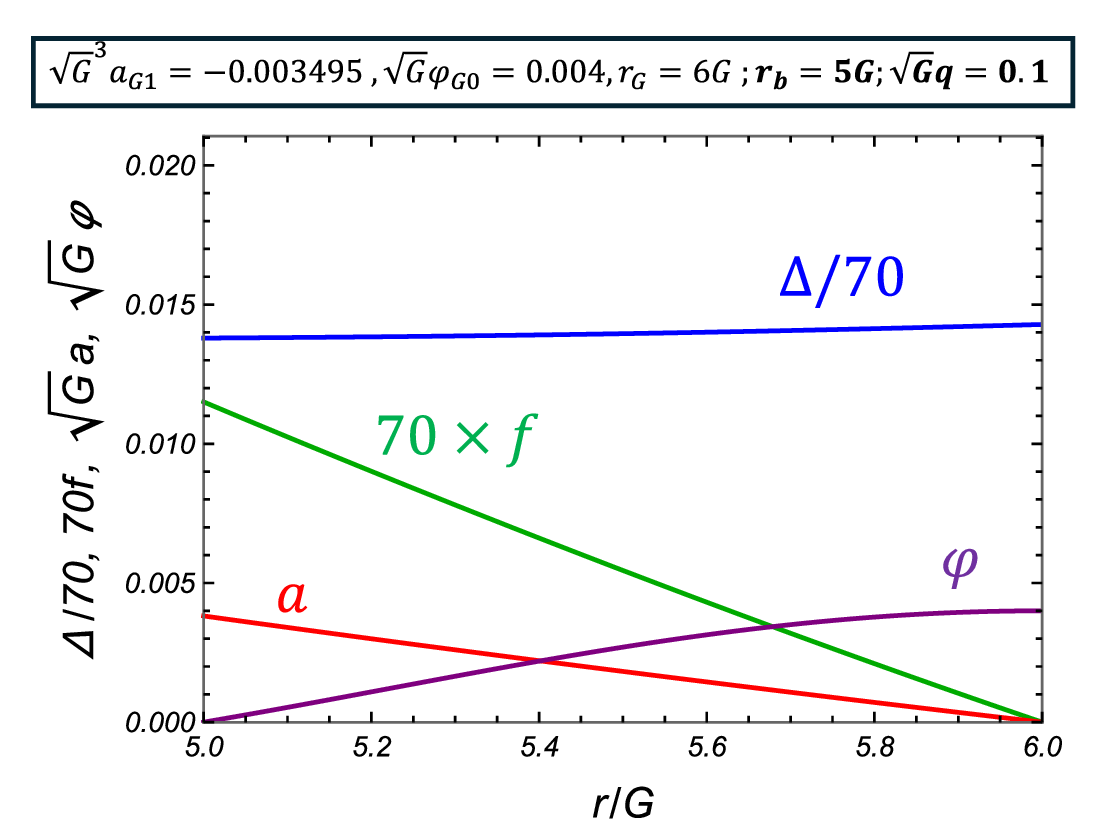} 
	\caption{\textbf{(left)}  Solution curve for the hairy BG on the space of $(a_{G1}, \varphi_{G0})$ with fixing $r_{G} = 6G$. The other parameters are set to $r_{b}=5G$ and $\sqrt{G}q = 0.1$. Since the $a_{G1}$ is negative and becomes exponentially small, I used $\log(-a_{G1})$ for the vertical axis.  \textbf{(right)} Example of a hairy BG solution. (I set $\sqrt{G}^3 a_{G1} = -0.003495, \sqrt{G}\varphi_{G0}=0.004, r_{G}=6 G, r_{b} = 5 G, \sqrt{G}q = 0.1$. }
\label{FIG7}
\end{center}
\end{figure}

In the above equation there are three parameters $(\Delta_{G0}, a_{G1}, \varphi_{G0})$, in addition to the horizon radius $r_{G}$. The first one $\Delta_{G0}$ can be set to 1, so there are three parameters in total, $(a_{G1}, \varphi_{G0})$ and $r_{G}$. Similar to the boson star(-PL) case, once we fix $r_{G}$, we have to shoot the appropriate points on the space of $(a_{G1}, \varphi_{G0})$ and they form a curve. A example is shown in Figure \ref{FIG7} \textbf{(left)}. 
                                                 %
                                                 %
As shown in this example, the corresponding $a_{G1}$ typically becomes exponentially small as the $\varphi_{G0}$ increases. Figure 8 \textbf{(right)} shows the behavior of the variables.
The qualitative behavior is the same as that in the upper hemisphere of the boson star-PL geometry. In the boson star-PL geometry, $\Delta$ and $\varphi$ monotonically decrease, while $f$ and $a$ increase, {\it from $r_{max}$ to $r_{b}$}. A similar behavior is observed in this hairy BG case {\it from $r_{G}$ to $r_{b}$}. 

Note that the left end of the curve $\varphi_{G0}\to 0$ in Figure 8 \textbf{(left)} corresponds to the hairless BG saddle which I introduced in eq. (\ref{EqSec3BG}) in the previous section, and the chemical potential $\mu$ is monotonically increased as we increase $\varphi_{G0}$. Therefore, there is an inequality for $\mu$ for a given $r_{G}$;
\bea
\sqrt{\frac{1}{2G} \log \frac{r_{G}}{r_{b}} } \leq \mu ~ ,
\ena
or in terms of $r_{G}$, it must satisfy the following inequality for a given $\mu$
\bea
r_{b} <r_{G} < r_{b}e^{2G \mu^2} ~ .
\ena
This is one of the remarkable properties of hairy BG in three dimensions,  since the horizon radius of the hairless BG in three dimensions (\ref{EqSec3BG}) is constant when we fix $\mu$ while that of hairy BG can change.

\subsection{Thermodynamics}

As I mentioned at the beginning of this section, there are a total of five types of saddles in this system. When $\mu$ is sufficiently small, only the empty saddle and the BG saddle are dominant. Then the behavior of the free energy versus temperature is the same as in the $q=0$ case (Figure \ref{FIG2} (\textbf{left})). When $\mu$ becomes high, the other saddles become dominant. Two examples of free energy behavior are shown in Figure \ref{FIG9}. The left figure (\textbf{left}) shows the case where the boson star saddle becomes dominant, i.e. I chose the parameter in the boson star region of Figure \ref{FIG5} (\textbf{right}). I also plotted the hairy BG branch in the figure. However, it is always subdominant in this case. The right figure (\textbf{right}) shows the case where the boson star-PL saddle and the hairy BG saddle become dominant. This time I have chosen the parameter in the boson star-PL region of Figure \ref{FIG5} (\textbf{right}). In this case, there are two phase transitions, one is the transition between boson star-PL and hairy BG, and the other is the transition between hairy BG and BG. 
                                                 %
\iffigure
\begin{figure}[t]
\begin{center}
	\includegraphics[width=7.cm]{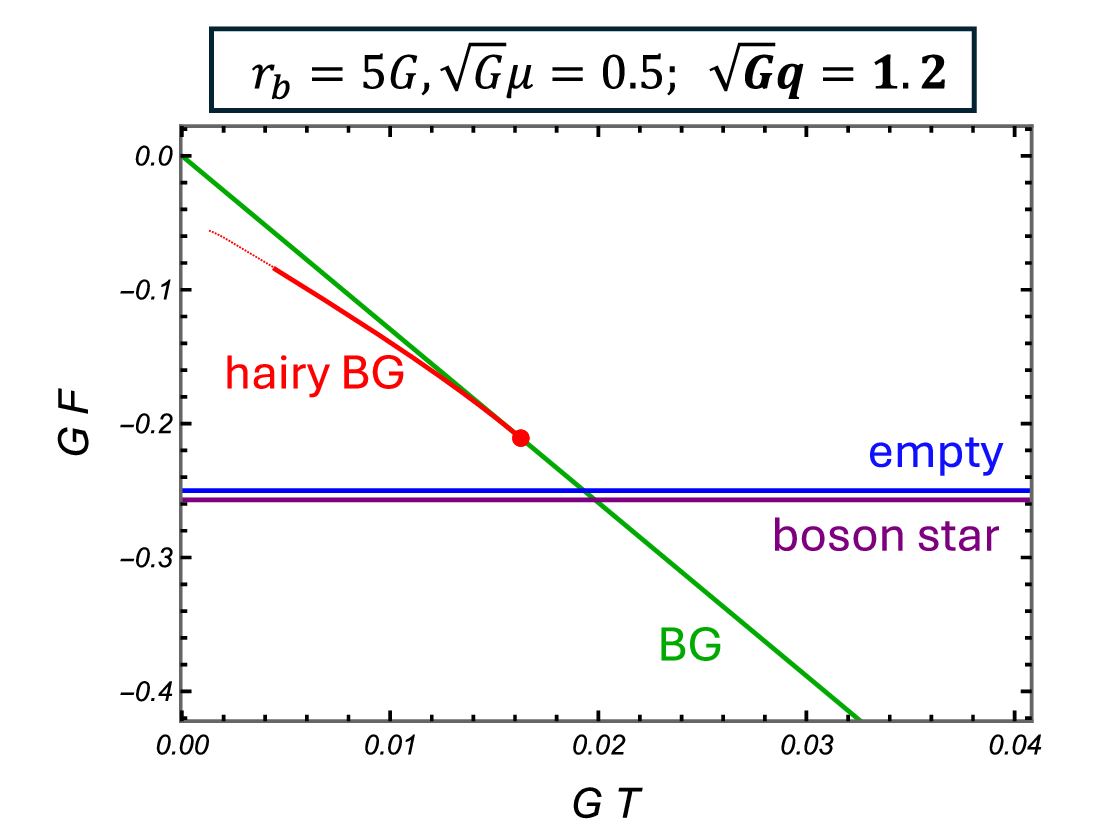} ~~ 	\includegraphics[width=7.cm]{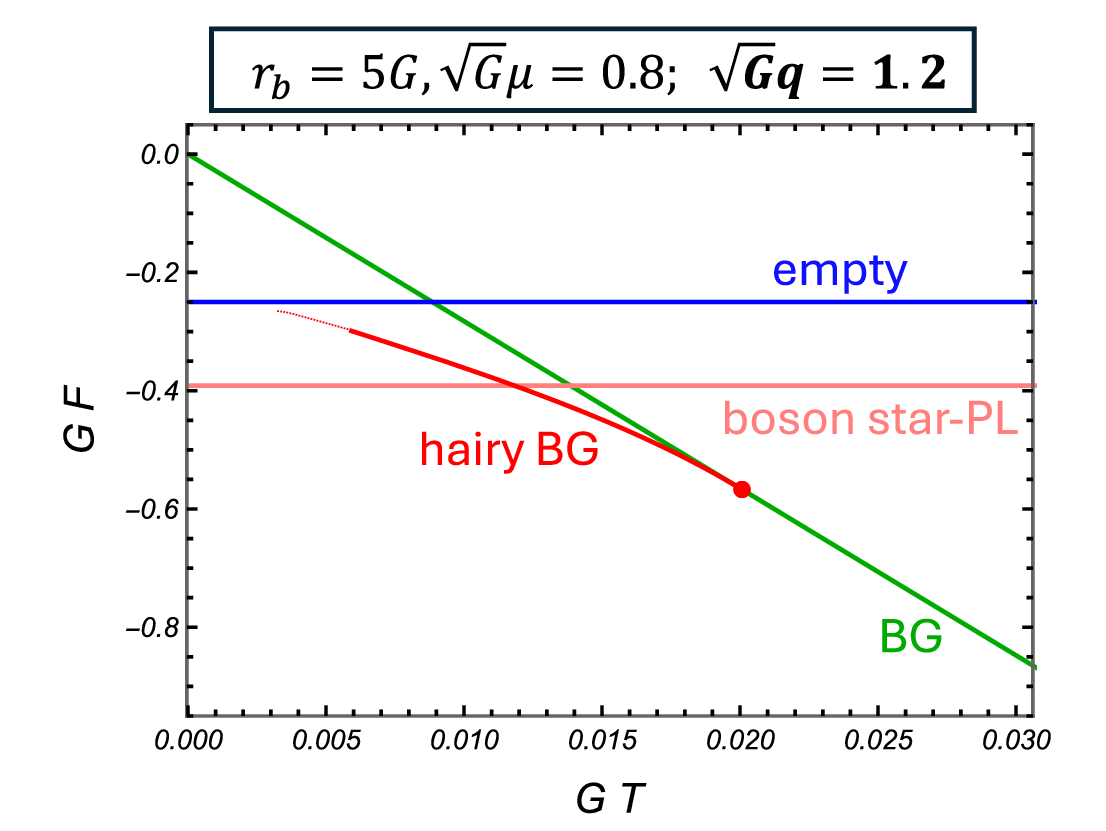} 
	\caption{Free energies versus temperature with fixed chemical potential. In both cases, I set $\sqrt{G}q =1.2$.  \textbf{(left)} The case of $\sqrt{G}\mu = 0.5$. In this case, the boson star saddle and the BG saddles are dominant. The empty saddles and hairy BG saddles are always subdominant. \textbf{(right)} The case of $\sqrt{G}\mu = 0.8$. The dominant saddles are the boson star-PL, the hairy BG, and the BG. \\ In both cases, I could not obtain the hairy BG solutions near zero temperature. This part corresponds to the dotted curve, and I do not know which direction it goes and where it ends. \\
The corresponding phase diagram is shown in Figure \ref{FIG10} (\textbf{left}).  }
\label{FIG9}
\end{center}
\end{figure}
\fi
                                                 %
Here, I will not try to show complete phase diagrams and their dependence of $q$. I only give two examples of phase diagrams with large $q$ and small $q$ in Figure \ref{FIG10}. 
                                                 %
\iffigure
\begin{figure}[h]
\begin{center}
	\includegraphics[width=7.5cm]{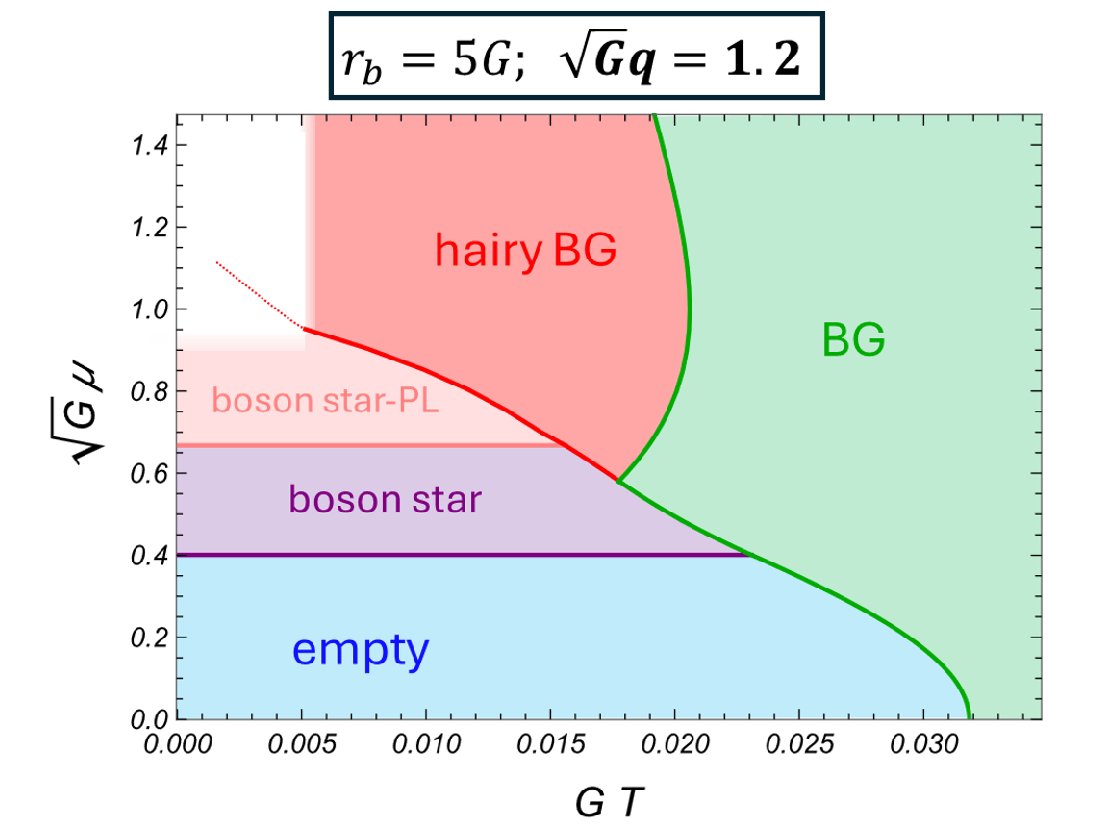} ~~ 	\includegraphics[width=7.5cm]{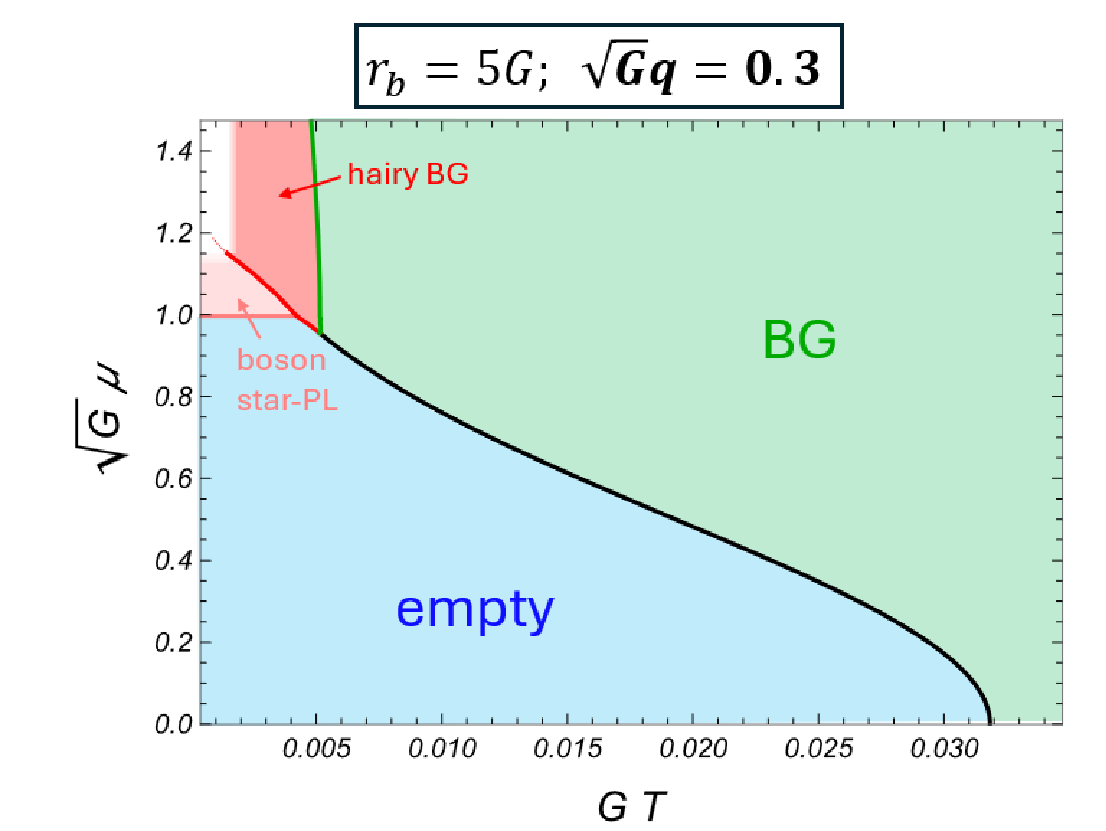} 
	\caption{\textbf{(left)}  The phase diagram of $\sqrt{G}q =1.2.$ The low $\mu$ and high $T$ regions are unchanged from the $q=0$ case (Figure \ref{FIG2} (\textbf{right})). Nontrivial phases appear in the high $\mu$ and low $T$ region. In this case, both the boson star phase and the boson star-PL phase appear.  \textbf{(left)} The phase diagram of $\sqrt{G}q =0.3$. The non-trivial region shrinks and the boson star phase is absent. The only solitonic phase is the unconventional one with a bulge as shown in Figure \ref{FIG4} \textbf{(right)}. \\
In both cases, I could not determine the boundary curve between the boson star-PL phase and the hairy BG phase in the very low $T$ region due to the same technical issue as in Figure \ref{FIG9}. }
\label{FIG10}
\end{center}
\end{figure}
\fi
                                                 %
Compared to the $q=0$ case (Figure \ref{FIG2} (\textbf{right})), the modification is only in the high $\mu$ and low $T$ region. As noted in the previous subsection, when $q$ is sufficiently large, both the boson star phase and the boson star-PL saddle exist as in \textbf{(left)} and when $q$ is low, the former phase is absent. We can see that the modified region shrinks as we decrease $q$. This would be consistent with the fact that we should recover the $q=0$ case when the limit $q\to 0$ is taken. If we consider the boson star phase and the boson star-PL phase as a single phase, the phase structure is similar to that in the systems of holographic superconductor \cite{NRT2010, HW2010} or \cite{BKB20161, AMS2016}.
\footnote{
Their setup is slightly different from the original holographic superconductor \cite{Gubser2008, HHH20081,MKF2010, HHH20082} which is the case of planar boundary. In \cite{NRT2010, HW2010}, they consider the boundary whose one direction is compactified to a circle. In \cite{BKB20161, AMS2016}, they considered the spherical boundary. 
}
If the similarity holds also for the very low $T$ region, where I could not obtain hairy BG solutions due to a technical problem, there could be ``zero temperature hairy BG'' when $q$ is sufficiently small. However, unlike the conventional case where the zero temperature limit is a zero horizon radius limit \cite{HR2009}, we may not be able to take a similar limit since the hairy solution is BG whose horizon must by definition be larger than the boundary radius $r_{b}$ in this case. Understanding the very low and zero temperature region is technically difficult, but worthy of careful analysis. I hope to return to this problem in the near future.

\clearpage
\section{Summary and Discussion}
In three dimensions without $\Lambda$, there is no BH solutions \cite{Ida2000, KSB2020}. Since in the familiar AdS setup BHs always play an important role in gravitational thermodynamics, one might worry that thermodynamics in three dimensions is rather boring. In the previous work \cite{Miyashita20242} and in the present paper, I have shown that this is not true. A new type of saddle, BG, which has a larger horizon area than the boundary area, plays the role of high temperature phase and the resulting phase structure is similar to and as complicated as the AdS case. In this paper, I investigated Einstein-Maxwell-scalar(EMS) system in three dimensions with Dirichlet boundary and found new types of saddles of this system. Thus, together with the previously known saddles, this system has five types of saddles; (i) empty saddle, (ii) BG saddle \cite{Miyashita20242}, (iii) boson star saddle \cite{KSB2020}, (iv) boson star-PL saddle, and (v) hairy BG saddle. As shown in Figure \ref{FIG10}, depending on $T$ and $\mu$ (and $q$), all of them appear as stable thermodynamic phases. Therefore, this three dimensional system still has many phases even though the BH phase is absent. In particular, in the AdS/CFT literature, the EMS system is known for the system of holographic superconductors \cite{HHH20082, Gubser2008, HHH20081, BBBLMU2010, BKB20161, AMS2016, NRT2010, HW2010} and the phase diagrams (Figure \ref{FIG10}) are very similar to those in \cite{NRT2010, HW2010} or \cite{BKB20161, AMS2016}. It would be very interesting to  investigate more finer quantities, such as conductivities, or to perform a detailed analysis of the phase diagram in order to see how this ``holographic superconductor'' differs from the conventional holographic superconductors.

In future work, it would be valuable to explore the low-temperature behavior of hairy BG saddles more carefully, possibly via improved numerical techniques or analytic methods. Moreover, generalizing these findings to higher-dimensional settings or coupling with additional fields could shed light on the universality and limits of the Dirichlet-boundary gravitational thermodynamics. Finally, exploring potential observational or holographic interpretations of the exotic saddles (such as the boson star-PL and hairy BG phases) remains an exciting direction.

The original purpose of this work is to construct a hairy version of BG saddles and to show that it also becomes thermodynamic phase. However, as a by-product, I also found that ``Python's lunch'' type solutions exist and become thermodynamic phase. It may be natural to expect that similar saddles exist and become dominant saddles in GPI for gravitational thermodynamics in higher dimensions or different systems. Therefore, we should be careful when analyzing gravitational thermodynamics with a Dirichlet boundary. The thermodynamic properties might be similar to the corresponding AdS boundary case, but there might be strange saddles in GPI. They could be bag of gold, python's lunch, or more exotic geometry.

\acknowledgemnet
This work is supported in part by the National Science and Technology Council (No. 111-2112-M259-016-MY3).

\appendix

\end{document}